\documentclass[preprint,authoryear,5p,11pt]{elsarticle}

\usepackage{amssymb,amsmath,amsfonts,epsfig,color,epstopdf}
\usepackage{algorithm,algorithmic}
\floatname{algorithm}{Algorithm}
\newcommand{\beq}{\begin{equation}}
\newcommand{\eeq}{\end{equation}}
\newcommand{\beqn}{\begin{eqnarray}}
\newcommand{\eeqn}{\end{eqnarray}}

\DeclareMathOperator*{\argmax}{arg\,max}
\def\bmath#1{\mbox{\boldmath$#1$}}

\usepackage{natbib}

\usepackage{makecell}
\usepackage{listings}
\usepackage{hyperref}

\usepackage{lineno}

\journal{Astronomy and Computing}

\begin{document}

\begin{frontmatter}



\title{Reinforcement learning}


\author{Sarod Yatawatta}

\affiliation{organization={ASTRON, The Netherlands Institute for Radio Astronomy},
            addressline={Oude hoogeveensedijk}, 
            city={Dwingeloo},
            country={The Netherlands}}

\begin{abstract}
  Observing celestial objects and advancing our scientific knowledge about them involves tedious planning, scheduling, data collection and data post-processing. Many of these operational aspects of astronomy are guided and executed by expert astronomers. Reinforcement learning is a mechanism where we (as humans and astronomers) can teach agents of artificial intelligence to perform some of these tedious tasks. In this paper, we will present a state of the art overview of reinforcement learning and how it can benefit astronomy.
\end{abstract}


\end{frontmatter}


\section{Introduction\label{intro}}
Reinforcement learning (RL), with the aid of advances in deep neural networks, has made major breakthroughs in diverse disciplines. Some early highlights were in computer games \citep{Atari}, in chess and Go \citep{silver2016mastering} and in robotics \citep{DDPG,SAC1}. Recent highlights include developing efficient algorithms such as in matrix multiplication \citep{fawzi2022discovering} and in sorting \citep{2023Natur.618..257M}.

There are a few applications of RL in astronomy as well. Telescope automation is closely related to robotics and RL can be used in telescope control including adaptive optics \citep{nousiainen2022toward,landman2021self,Nousiainen21} and adaptive reflective surface control \citep{peng2022intelligent} as well as in observation scheduling \citep{jia2023observation,jia2023simulation,jia2022optimal}. Going further down the data flow, RL has been applied in radio astronomical data processing pipelines \citep{Y2021,yatawatta2023hint} for hyper-parameter tuning. Considering modern astronomy to be a flow of data or information from the observing telescope to the scientist, we foresee many more applications of RL to aid and refine this flow and motivates this publication.

Several methodologies fall under the umbrella of machine learning (ML): Supervised learning is the most commonly used methodology where a machine is given both the input and the required output to learn to perform a certain task. In unsupervised learning on the other hand, only the input is given to the machine. Reinforcement learning follows a different paradigm where a machine learns to perform a task by repeated attempts and getting some form of feedback from an external environment. Another noteworthy difference in RL is the temporal aspect, i.e., the task to perform is considered to be a sequence of actions to take instead of just one action, as in, say, supervised learning where a classifier outputs the class corresponding to the input in one step.

In this paper, we provide an overview of modern deep-RL with a focus on its use in astronomy. Reinforcement learning has a long history and multiple origins, stemming from several disciplines such as machine learning, dynamic programming and control and the scope of of this paper is to give a brief but sufficient overview of the topic so that a new user can quickly apply the RL techniques in their work. This paper is organized as follows: in section \ref{theory}, we give a theoretical overview of RL. In section \ref{algo}, we discuss model-free RL algorithms both for discrete and continuous action spaces. Next in  section \ref{modelbased}, we discuss model-based RL where a model representing the environment is built and used. Finally, in  section \ref{app} we discuss practical aspects of RL including commonly used software and we conclude in section \ref{conc}.

Notation: We do not distinguish between scalars, vectors or tensors in our notation and the actual dimension of the variables should be inferred depending on the context of their use. The matrix transpose, inverse and determinant are given as $(\cdot)^T$, $(\cdot)^{-1}$ and $\det(\cdot)$, respectively. A multinomial Gaussian with mean $\bmath{\mu}$ and covariance $\Sigma$ is given as $\mathcal{N}(\bmath{\mu},\Sigma)$.

\section{Reinforcement learning theory\label{theory}}
In this section, we provide a concise theoretical overview of reinforcement learning . Modern RL sits in the intersection of many diverse disciplines and we take an approach based on machine learning \citep{SuttonBarto,2010Szepesvari} to provide this introduction. Alternative viewpoints do exist, for example from dynamic programming and control \citep{bertsekas2005dynamic,bertsekas2012dynamic} or even from neuroscience \citep{10.5555560669}.
\subsection{The state, action and reward}
\begin{figure}[ht]
\begin{minipage}{0.98\linewidth}
\begin{center}
\epsfig{figure=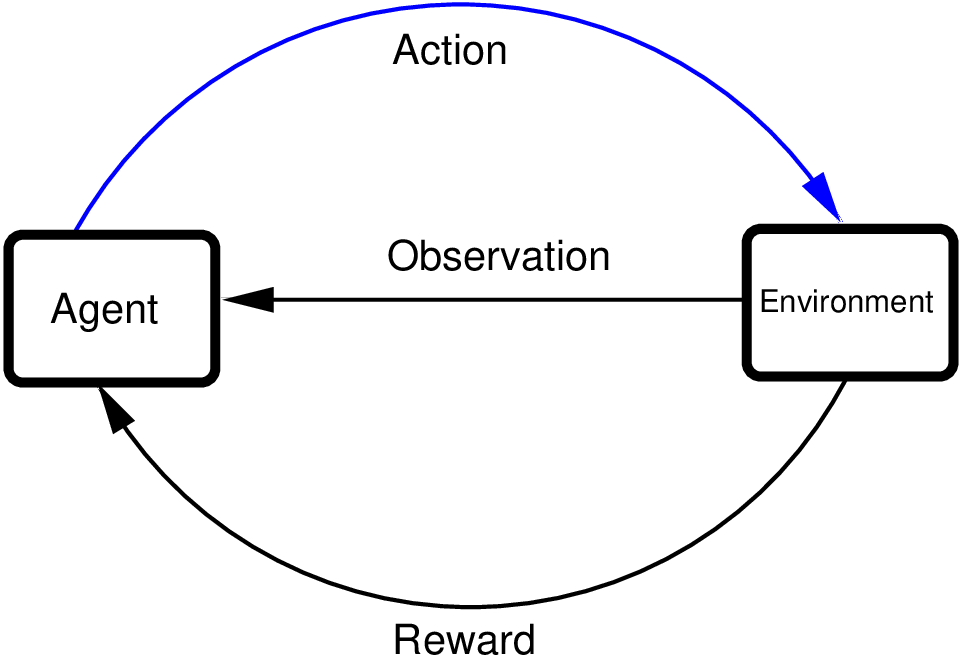,width=8.0cm}\\
\end{center}
\end{minipage}
  \caption{An agent interacting with its environment. The agent receives an observation and performs an action and receives a reward corresponding to the action.\label{agent_env}}
\end{figure}

We consider a form of artificial intelligence, which we call {\em the agent} interacting with its {\em environment} as seen in Fig. \ref{agent_env}. The agent performs a task and has an objective or a goal that we can specify. Information is passed from the environment or the world to the agent in the form of {\em observations}. After considering the observations, the agent performs the task which is fed to the environment as the {\em action}. Based on the action, the environment will undergo a change. To evaluate the effect of the action taken, the agent also receives a {\em reward} which is a numeric evaluation of the quality of the action and its effect on the environment.

We put the relationships between the agent and its environment into a more rigorous mathematical form as follows. Let $\mathcal{S}$ be a set whose elements are possible representations of the {\em state} of the environment. Any element in $\mathcal{S}$ can be given by $s$ and without loss of generality we assume $s$ to be a vector. Note that the state does not have to be equal to the observation: the state is a condensed form of the observation, for example by removing observables that are directly dependent on other observables. We consider $\mathcal{A}$ to be a set of possible actions to take. Any element in $\mathcal{A}$ can be given by $a$ which is also assumed to be a vector. Note also that both $s$ and $a$ can be vectors of discrete values (integers) as well as continuous values depending on the problem. In Table \ref{envs} we have listed several examples of RL applications. In the chess game example, both the state and action are discrete because there are only a finite number of chess pieces and moves that can be made at any point in the game. In the other two examples, we have continuous values for $s$ and $a$ although some state (or action) variables can be discrete. The dimensions of $s$ and $a$ can also be large for real life problems, for example in the self-driving automobile example in Table \ref{envs}. 
\begin{table*}[htbp]
\begin{minipage}{0.98\linewidth}
\caption{Some examples where RL can be applied} \label{envs}
  \begin{center}
   \begin{tabular}{clll}
     \thead{Problem} & \thead{Objective} & \thead{State} & \thead{Action} \\\hline \\
     Chess & Win game& \makecell{Positions of chess\\ pieces on the board.} & \makecell{Select and\\ move a chess piece.}\\
     \makecell{Walking robot\\ (Bipedal walker)}  & Walk& \makecell{Positions, velocities\\ of leg joints etc.} & \makecell{Apply torque to\\ various leg joints.}\\
     \makecell{Self driving\\ automobile}  & Reach destination& \makecell{Position, velocity,\\ acceleration of self\\ and other vehicles;\\ fuel level, road measurements, road signs,\\ pedestrians etc.} & \makecell{Apply forces\\ for acceleration, braking,\\ steering etc.}\\
   \end{tabular}
  \end{center}
\end{minipage}
\end{table*}

The quality of the action taken by the agent is measured by the {\em reward} $r$ which is generally a real valued scalar. We consider the function generating the reward to be in the functional space $\mathcal{R}$. The definition of the reward is unique to each RL problem and we (as the users) have the freedom to define this as long as we follow the general rule: the higher the reward is, the closer we are to achieving our objective. In other words, we give higher rewards to the agents when the actions taken by the environment leads to reaching the goal and lower rewards (or higher penalties) when the actions lead to poor results or failure.

\subsection{Markov decision processes}
As seen in Fig. \ref{agent_env}, the interactions between the agent and its environment are repetitive, i.e., at any given iteration, the agent observes and receives the state $s$ (and reward $r$) and recommends action $a$ which is passed onto the environment. The environment takes the action $a$ and as a result, its state will change. This updated state (or observation) is passed onto the agent, together with the reward $r$. This cycle continues until we reach our goal, hence it is a {\em process}. Most RL problems are {\em episodic}, i.e., after many repetitions of this cycle, the agent reaches its goal or ends in failure. If we consider the chess game example in Table \ref{envs}, the game ends when the agent wins, loses or draws the chess game. Therefore, one chess game can be considered as one episode. In the bipedal walker and self-driving automobile examples, the episode will end when we reach our destination, or when we run out of energy (fuel) or when we meet with an accident. This temporal aspect is unique to RL compared with other ML problems such as supervised learning. We use the subscript $t$ to denote this temporal dependence when we describe the state, action and the reward, such as $s_t$, $a_t$ and $r(s_t,a_t)$. At step $t$, the transition probability of the state from $s_t$ to the next state $s_{t+1}$ is given by $p$ which is called the {\em state transition probability} in $\mathcal{P}$.

The tuple $(\mathcal{S},\mathcal{A},\mathcal{R},\mathcal{P})$ is called a Markov decision process (MDP). The Markov property implies that the state transition probability at step $t$ is only dependent on the current state $s_t$ and the action taken $a_t$. In other words, we can parameterize the state transition probability as $p(s_{t+1}|s_t,a_t)$ which is only dependent on $s_t$ and $a_t$ (we use the notation for conditional probability here). The reward received at state $s_t$ after performing action $a_t$ is denoted by $r(s_t,a_t)$ which can also be represented as $r_{t}$.

An important question to answer in many RL problems is how to determine if the agent has learnt to perform the required task. Considering the examples in Table \ref{envs}, it is clear for the chess example where we consider the agent has learnt to play chess if it wins in almost all games. For the bipedal walker example, determining whether or not the walker has learnt to walk is a bit problematic. In order to alleviate this issue we consider the agent has learnt the task if it can sustain a sufficiently high reward at each time step $t$. Obviously, we do not expect the agent to reach such a high reward at start and we need to look forward to the future. Therefore, while learning a task, the agent does not attempt to maximize the {\em immediate} reward, but the cumulative reward over a number of future steps as well. In an {\em infinite horizon} MDP, we consider an infinite number of steps in the future. To account for uncertainties in the future, we calculate a {\em discounted} cumulative reward with a discount factor $\gamma$ ($0<\gamma<1$). This also makes the summation of rewards over future steps converge to a finite value.

\subsection{Q function, value function and policy}
\begin{figure}[ht]
\begin{minipage}{0.98\linewidth}
\begin{center}
\epsfig{figure=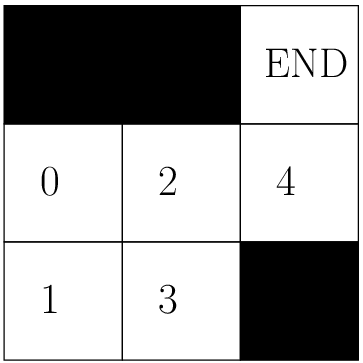,width=5.0cm}\\
\end{center}
\end{minipage}
  \caption{The maze environment with $5$ valid states $0,1,\ldots,4$. The agent can move (act) $\leftarrow$,$\rightarrow$,$\uparrow$, or $\downarrow$. The state $\mathcal{S}$ is a discrete space with $5$ states and the action $\mathcal{A}$ is also a discrete space with $4$ actions.\label{maze}}
\end{figure}

In order to illustrate the basic RL concepts, we start with a simple example: In Fig. \ref{maze} we show a maze, which is our environment. The agent can start from any of the empty squares and has to navigate to the top right hand square, which is the goal. At each step, the agent can make four moves, i.e., up, down, left and right. If the agent reaches the top right hand corner, we reach the end of the episode (terminal state).

In order to reach our goal, we can define the reward given for each action taken at each state as in Table \ref{maze_r}. Note that we give a very low reward ($-\infty$) to impossible actions. We give a high reward ($100$) when we reach our goal by moving up at state $4$.

\begin{table}[htbp]
\begin{minipage}{0.98\linewidth}
  \caption{Tabulated reward $R[s,a]$ for the maze environment to achieve the goal. Each row corresponds to a state and each column corresponds to an action.} \label{maze_r}
  \begin{center}
   \begin{tabular}{l|cccc}
      & $\rightarrow$ & $\leftarrow$ & $\uparrow$ & $\downarrow$\\\hline
     $0$  & $-1$ & $-\infty$ & $-\infty$ & $-1$\\
     $1$  & $-1$ & $-\infty$ &  $-1$ & $-\infty$\\
     $2$  & $-1$ & $-1$ &  $-\infty$ &  $-1$\\
     $3$ & $-\infty$ & $-1$ & $-1$ &  $-\infty$ \\
     $4$ & $-\infty$ & $-1$ & $100$ & $-\infty$ \\
   \end{tabular}
  \end{center}
\end{minipage}
\end{table}

In order to reach our goal in the maze, we tabulate the {\em quality} of each state and action pair that we name as the {\em Q-table}. The Q-table therefore is a table with rows corresponding to each state and columns corresponding to each action (similar form as Table \ref{maze_r}). Let us call this $Q[s,a]$ and let us call the reward table in Table \ref{maze_r} as $R[s,a]$. We follow algorithm \ref{qtab_iter} to update the values of $Q[s,a]$ iteratively.

\begin{algorithm}
\caption{Q-table iteration \label{qtab_iter}}
  \begin{algorithmic}[1]
    \REQUIRE Discount $\gamma$, Reward table $R$
    \STATE Initialize Q-table to all zeros.
    \STATE Select random initial state $s$ from $\{0,1,\ldots,4\}$.
    \WHILE{not reached terminal state}
      \STATE Select action $a$ $ \leftarrow$ column number that gives maximum value of row number $s$ in Q-table. If more than one column has the maximum value, choose randomly between them.
      \STATE Next state $s^\prime \leftarrow$ follow action $a$ and Fig. \ref{maze}.
      \IF {$s^\prime=$ terminal state}
        \STATE Update $Q[s,a] \leftarrow R[s,a]$.
        \STATE Stop.
      \ELSE
      \STATE Update $Q[s,a] \leftarrow R[s,a]+\gamma($ maximum value of row $s^\prime$ in Q-table $)$.
      \STATE Next state $s \leftarrow s^\prime$.
      \ENDIF
    \ENDWHILE
  \end{algorithmic}
\end{algorithm}

What we have shown in algorithm \ref{qtab_iter} is a crude form of {\em Q-learning} that is only feasible with a low dimensional, discrete state and action spaces. The application of algorithm \ref{qtab_iter} to Fig. \ref{maze} is tedious, but can be done by hand with some value selected for $\gamma$ (say $\gamma=0.9$). Python source code implementing algorithm \ref{qtab_iter} for the maze environment is given in \ref{q-python}. After sufficient number of iterations and sufficient repetitions of algorithm \ref{qtab_iter} with different initial states (episodes), we will get an end result as in Table \ref{maze_qtable}. 

\begin{table}[htbp]
\begin{minipage}{0.98\linewidth}
\caption{Converged Q-table for the maze environment after several iterations with $\gamma=0.9$.} \label{maze_qtable}
  \begin{center}
   \begin{tabular}{l|cccc}
      & $\rightarrow$ & $\leftarrow$ & $\uparrow$ & $\downarrow$\\\hline
     $0$  & $79.1$ & $0$ & $0$ & $62.171$\\
     $1$  & $70.19$ & $0$ &  $70.19$ & $0$\\
     $2$  & $89$ & $70.19$ &  $0$ &  $70.19$\\
     $3$ & $0$ & $62.171$ & $79.1$ &  $0$ \\
     $4$ & $0$ & $0$ & $100$ & $0$ \\
   \end{tabular}
  \end{center}
\end{minipage}
\end{table}

Using Table \ref{maze_qtable}, we can solve the maze environment, for example in state $0$, we get the highest Q-value ($79.1$) by taking action in the first column $\rightarrow$ while in state $1$, we can choose either the first $\rightarrow$ or the third $\uparrow$ column and so on. In most problems however, the dimensions of the state space and the action space are too high to use a tabular method. The same can be said if the state or action spaces are continuous. 

In order to generalize RL algorithms to handle high dimensional or continuous problems, we move on from a tabular representation to a functional representation relating $\mathcal{S}$ and $\mathcal{A}$. Some of the major components are:
\begin{itemize}
  \item Policy: The policy is a mapping from $\mathcal{S}$ to $\mathcal{A}$. A deterministic policy $\pi(s)$ will produce action $a$ given the state $s$, $\pi(s)\rightarrow a$. A stochastic policy $\pi(a|s)$ will predict the conditional probability distribution of the action $a$ given the state $s$. We can generate samples from the conditional probability distribution to get a representation of the action, i.e., $a\sim \pi(a|s)$.
  \item Q-function: The Q-function $Q(s,a)\rightarrow q$ is a mapping from the state and action spaces to a real number $q$, $\mathcal{S}\times \mathcal{A} \rightarrow \mathbb{R}$. This is the generalization of the Q-table (e.g., Table \ref{maze_qtable}) from discrete RL problems to continuous or high dimensional RL problems. Given the state $s$, $Q(s,a)$ will given an indication of the quality (or expected cumulative reward) of taking action $a$ under policy $\pi(\cdot)$. The higher $Q(s,a)$ is, the closer we are to achieving the goal of the RL problem. The Q-function is also called action-value function or state-action value function.
  \item Value: The value of state $s$ is the expected cumulative reward starting with state $s$ and following a policy $\pi(\cdot)$. In other words, it is a measure of the importance of being in state $s$ to reach the goal of the RL problem. We use $V(s)\rightarrow v$ to denote the value function that maps $\mathcal{S}\rightarrow \mathbb{R}$.
\end{itemize}

It is important to understand how the aforementioned policy and value functions are related to one another. At the solution of the RL problem (or after reaching the goal), the optimality conditions are defined by the Bellman equation (\ref{bellman})
\beq\label{bellman}
Q(s,a)= r(s,a) + \gamma \underset{a^\prime=\pi(s^\prime)}{\mathrm {max}} Q(s^\prime,a^\prime).
\eeq
The Bellman equation relates $Q(s,a)$ to the (optimal) policy $\pi(s)$. The current state and action taken are given by $s$ and $a$ respectively while the next state and action (under optimal policy) are given by $s^\prime$ and $a^\prime$. The immediate reward is given by $r(s,a)$ while the discount factor is $\gamma$. Note that we have already used a form of (\ref{bellman}) in algorithm \ref{qtab_iter} to solve the maze problem (see line 10). Note also that if $s$ is the terminal state (end of episode), $s^\prime$ does not exist and the right hand of (\ref{bellman}) is just $r(s,a)$.

To solve any given RL problem, we have to learn the optimal $\pi(s)$, $Q(s,a)$ or $V(s)$ tailored to that particular problem. In order to do this, we represent each of the aforementioned functions as deep neural networks (DNN, \cite{LeCun2015}). There are two main reasons for deep neural networks to be ideally suited for this task. First, deep neural networks have the ability to find arbitrary representations. Second, from a practical viewpoint, modern deep learning frameworks have built in learning capabilities with gradient descent and built-in gradient calculation such as reverse mode automatic differentiation \citep{paszke}. The deep neural networks are parameterized by trainable parameters and in section \ref{algo} we will describe how we train these deep neural networks to solve the RL problem.

\section{Deep reinforcement learning algorithms\label{algo}}
In this section, we will focus our attention on model-free deep RL algorithms. We first discuss the common difficulties faced during training an RL agent.
\begin{itemize}
  \item Not enough data: A large amount of training data is required even in other deep learning tasks such as supervised learning. This is even more prominent in RL. Note that in RL, we generate training data by interacting with an environment. Most real life environments are complex and expensive to operate (for example a self driving automobile). Hence, generating enough training data is hard in RL. In section \ref{modelbased}, we discuss model-based RL as a way to generate data without interacting with an environment. Alternatively, we can store and re-use past experience which we will elaborate in this section.
  \item Exploitation vs exploration: The optimization problems underlying the training of RL are non-convex and ill-conditioned with high dimensional parameter spaces. Moreover, the state space $\mathcal{S}$ and action space $\mathcal{A}$ can also be high dimensional. In order to reach the global optimal point in our optimization, we need to uniformly sample the full $\mathcal{S}$ and $\mathcal{A}$. It is easy for the solutions to converge to local minima or overfit because of poor sampling. In order to overcome this, during selection of the action to take, we can balance between random sampling (exploration) or using the policy to maximize the reward (exploitation). This is commonly termed $\epsilon$-greedy action selection, where we select a random action with probability $\epsilon$ and we exploit the policy with probability $1-\epsilon$ where $\epsilon$ is a small positive value.
  \item Stability: The solution to the RL problem is obtained by (indirectly) solving (\ref{bellman}) or something similar in an iterative manner. We do see that $Q(s,a)$ appears on both sides of the equation and makes in unstable. Generally a small learning rate is chosen to avoid divergence but other improvements are used to overcome this as we illustrate later in this section.
\end{itemize}
\subsection{Experience replay}
In order to alleviate the data deficiency for training, we can store our interactions with the environment at each step $t$ and re-use them as {\em experience}. We keep a special buffer (replay buffer) $\mathcal{D}$ for this and at each step, we store the tuple $($ state $s$, action $a$, reward $r$, next state $s^\prime$ $)$ in the buffer. Since we are using past experience, we are using actions based on past policies, that might be different from the current policy. Hence the use of a replay buffer compels us to use {\em off-policy} RL algorithms. It is also possible to prioritize the past experience that we retrieve from $\mathcal{D}$ based on how poor the agent has performed in the past \citep{prioritized_experience}. For example, we can prioritize retrievals from $\mathcal{D}$ giving high priority to past steps where the transitions from $s$ to $s^\prime$ has large change in the Q-values (importance sampling).

We present the high level algorithm for RL using a replay buffer in algorithm \ref{algRL}. This algorithm uses transitions from step $t$ to $t+1$ for learning, also called temporal difference learning with one step look ahead (TD-0). We use two outer loops, one loop over $E$ episodes and an inner loop over $L$ steps per-episode. We consider an infinite-horizon RL problem and hence make $L$ sufficiently large for this purpose.

\begin{algorithm}
\caption{Training the RL agent}
\label{algRL}
\begin{algorithmic}[1]
\REQUIRE Number of episodes $E$, number of loop iterations $L$
\STATE Initialize Environment and Agent.
\STATE Setup empty (or reload saved) ReplayBuffer $\mathcal{D}$.
\FOR{$e=1,\ldots,E$}
\STATE Environment: simulate or realize new (random) observation.
\STATE Agent: $\leftarrow$ initial state $s$.
\FOR{$t=1,\ldots,L$}
 \STATE $a \leftarrow$ Agent suggests action based on $s$.
  \STATE Environment: take action $a$, determine reward $r(s,a)$ and next state $s^\prime$.
  \STATE $\mathcal{D}$: store $(s,a,r,s^\prime)$.
  \STATE Agent: sample a mini-batch from $\mathcal{D}$ and learn.
  \IF {$s^\prime$ is not terminal state}
   \STATE $s\leftarrow s^\prime$
  \ELSE
  \STATE break loop
  \ENDIF
\ENDFOR
\ENDFOR
\end{algorithmic}
\end{algorithm}

An important statement in algorithm \ref{algRL} is in line 10, where the actual training of the agent is done. A {\em mini-batch} is one or more tuples of $(s,a,r,s^\prime)$ that is sampled from $\mathcal{D}$ and fed to the learning algorithm. Generally, a large minibatch will give a more stable performance of the learning algorithms and the largest size of the mini-batch is mostly determined by the available memory. We will expand the learning statement (line 10 in algorithm \ref{algRL}) in the following to consider various learning strategies.

\subsection{Discrete action RL}
In a discrete action space $\mathcal{A}$, the action $a$ can only take a finite number of values. However the state space $\mathcal{S}$ can be continuous or discrete, we do not impose any restriction. In contrast to the tabular algorithm \ref{qtab_iter}, we use a Q-function $Q(s,a)$ where $a$ is discrete. The Q-learning \citep{watkins1992q} step can be given as in (\ref{q-learning}): 
\beqn \label{q-learning}
\lefteqn{Q(s,a)\leftarrow Q(s,a)}\\\nonumber
&&+\mu\left(r(s,a)+\gamma \underset{a^\prime}{\mathrm {max}} Q(s^\prime,a^\prime) - Q(s,a)\right).
\eeqn
Note that we have not (yet) discussed how $Q(s,a)$ is represented in (\ref{q-learning}), this can be done by a table (with interpolation for $s$) or by a deep neural network. The learning rate in (\ref{q-learning}) is given by $\mu$. Because $a$ is discrete and has only a finite number of values, evaluation of $\underset{a^\prime}{\mathrm {max}} Q(s^\prime,a^\prime)$ can be done by evaluation of $Q(s^\prime,a)$ for all possible values of $a$ and selecting the maximum Q-value. If we have more than one tuple of $(s,a,r,s^\prime)$ in a mini-batch, we repeat (\ref{q-learning}) for all tuples sequentially. Initial values for $Q(s,a)$ will be set to $0$, especially for the terminal states $s$. If the next state $s^\prime$ is terminal, the right hand term of (\ref{q-learning}) is just $\mu \left(r(s,a) - Q(s,a)\right)$ and the same is implicitly applied to all learning steps hereafter.

As noted before, solving (\ref{bellman}) by iterating (\ref{q-learning}) is not stable and leads to overestimation \citep{van2016deep}, especially for large $\mu$ values. Double Q-learning \citep{hasselt2010double} is an improvement to Q-learning in this regard. In double Q-learning, instead of one, we use two Q-functions, i.e., $Q_1(s,a)$ and $Q_2(s,a)$. The motivation for this is to keep one Q-function fixed while updating the other. With probability $0.5$ we update the first Q-function as
\beqn \label{dq-learning-1}
\lefteqn{Q_1(s,a)\leftarrow Q_1(s,a)}\\\nonumber
&&+\mu\left(r(s,a)+\gamma  Q_2\left(s^\prime, \underset{a^\prime}{\argmax} Q_1(s^\prime,a^\prime)\right)\right.\\\nonumber
&&\left.- Q_1(s,a)\right)\\\nonumber
\eeqn
and otherwise we update the second Q-function
\beqn \label{dq-learning-2}
\lefteqn{Q_2(s,a)\leftarrow Q_2(s,a)}\\\nonumber
&&+\mu\left(r(s,a)+\gamma  Q_1\left(s^\prime, \underset{a^\prime}{\argmax} Q_2(s^\prime,a^\prime)\right)\right.\\\nonumber
&&\left.- Q_2(s,a)\right).\\\nonumber
\eeqn
Note that in (\ref{dq-learning-1}) and (\ref{dq-learning-2}), the target reward is estimated by the Q-function which is fixed. The evaluation of $Q_2\left(s^\prime, \underset{a^\prime}{\argmax} Q_1(s^\prime,a^\prime)\right)$ can be described as first finding the value $a^\prime$ that gives the maximum of $Q_1(s^\prime,a^\prime)$ and using this to evaluate $Q_2(s^\prime,a^\prime)$.

Thus far, we have treated the Q-functions without any consideration on how they are represented. In most practical algorithms, the Q-functions are represented by deep neural networks. For example, using a DNN with trainable weights given by $\theta$, we can model the Q-function in (\ref{q-learning}) and we denote this by $Q_\theta(s,a)$. Using the same trick of double Q-learning, we create two Q-functions, one parameterized by $\theta$ which is trained and another parameterized by $\theta^\prime$ which is denoted as the target Q-function, i.e., $Q_\theta(s,a)$ and $Q_{\theta^\prime}(s,a)$ respectively.

We minimize the mean squared error loss $J(\theta)$
\beq \label{Jloss}
J(\theta)=\|r(s,a)+\gamma \underset{a^\prime}{\mathrm {max}} Q_{\theta^\prime}(s^\prime,a^\prime) - Q_\theta(s,a)\|^2
\eeq
during training. A gradient descent step to minimize (\ref{Jloss}) can be given as
\beq \label{thetaup}
\theta \leftarrow \theta - \mu \nabla_\theta J(\theta)
\eeq
where $\mu$ is the learning rate. Note that the gradient of (\ref{Jloss}) can be calculated as
\beqn \label{Jgrad}
\lefteqn{\nabla_\theta J(\theta)=}\\\nonumber
&&-2\left(r(s,a)+\gamma \underset{a^\prime}{\mathrm {max}} Q_{\theta^\prime}(s^\prime,a^\prime) - Q_\theta(s,a)\right)\\\nonumber
&&\times \nabla_\theta Q_\theta(s,a)
\eeqn
using the chain rule. Note the similarity of the gradient (\ref{Jgrad}) to the Q-learning step (\ref{q-learning}). The target network parameters $\theta^\prime$ are updated at a lower cadence after repeating (\ref{thetaup}) several times by copying $\theta$, i.e., $\theta^\prime \leftarrow \theta$. For a mini-batch of several $(s,a,r,s^\prime)$, the loss (\ref{Jloss}) is averaged over the whole mini-batch. With modern deep learning frameworks, we only need to specify the loss to minimize as in (\ref{Jloss}) and it is not necessary to explicitly calculate the gradient and the gradient descent.

Looking back at Table \ref{envs}, the chess game example has both discrete state and action spaces. However, their dimensionalities are high, and unlike other RL problems, the environment (another chess player) acts in an adversarial manner. For this type of RL problems (games) a combination of Monte Carlo tree search (for dimensionality reduction) and DNNs (to reduce depth and breadth of tree search using value functions) are used \citep{silver2016mastering}.
\subsection{Continuous action RL}
In a continuous action space, the action $a$ can have an infinite number of values. Therefore, the direct application of algorithms such as Q-learning is not feasible. The policy $\pi(s)$ or $\pi(a|s)$ plays a major role in calculating $a$, instead of searching through all possible actions. In dynamic programming, there are two families of algorithms for solving RL problems in continuous action spaces: value iteration and policy iteration. In value iteration, the value function $V(s)$ is updated iteratively (according to the Bellman optimality condition (\ref{bellman})) until convergence and based on the converged value function, the policy is calculated. On the other hand, in policy iteration, the value function is updated while evaluating the current policy and thereafter, the policy is also iteratively updated.

\begin{figure}[ht]
\begin{minipage}{0.98\linewidth}
\begin{center}
\epsfig{figure=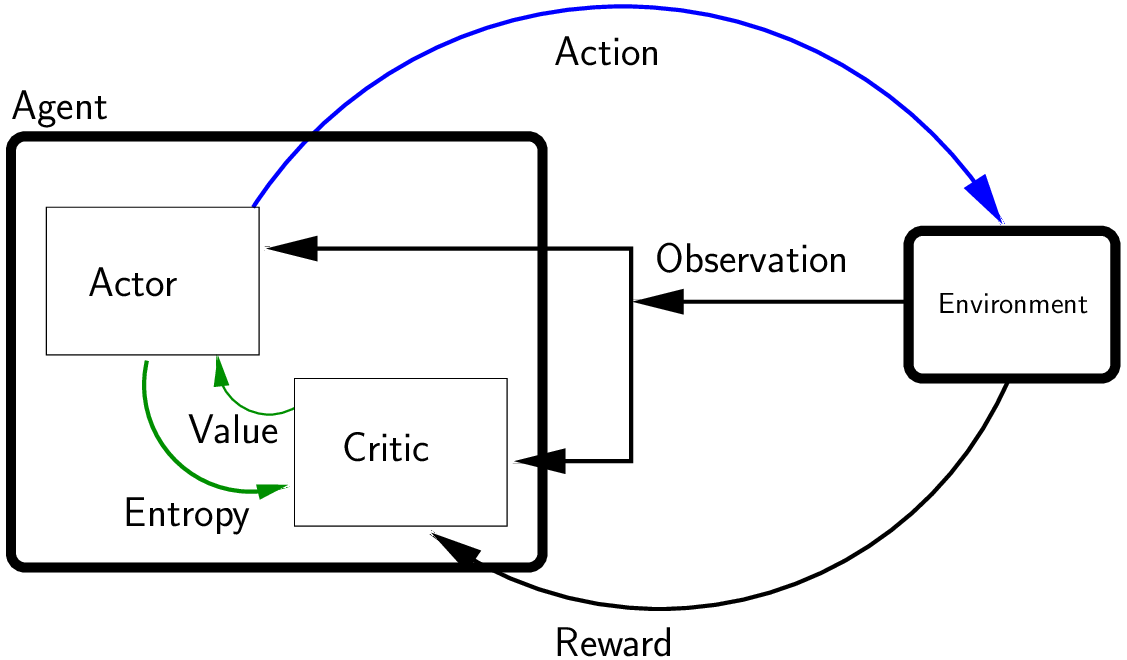,width=8.0cm}\\
\end{center}
\end{minipage}
  \caption{An RL agent composed of an actor and a critic.\label{actor_critic}}
\end{figure}

In this paper however, we focus on actor-critic methods where we jointly update both the value function and the policy and can be considered as a combination of value iteration and policy iteration. As seen in Fig. \ref{actor_critic}, we decompose the agent into an actor and a critic as follows:
\begin{itemize}
  \item Actor: implements the policy. The actor will produce and action $a$ given state $s$, $\pi(s)\rightarrow a$ or the conditional probability of $a$ given state $s$, $\pi(a|s)$ that can be sampled to generate $a$. In algorithm \ref{algRL}, the actor is active in line 7 and line 10. 
  \item Critic: evaluates the state $s$ (and action $a$) using the Q-function $Q(s,a)$ or the value function $V(s)$. Conceptually, the critic provides a critique of the action taken by the actor. The critic is active in the learning step in line 10 of algorithm \ref{algRL}.
\end{itemize}

We describe some of the actor critic algorithms that are state-of-the-art in the following.
\subsubsection{Deep deterministic policy gradient (DDPG)}
In DDPG \citep{DDPG}, the actor uses a deterministic policy parameterized by parameters $\phi$, i.e., $\pi_\phi(s)$. The critic is parameterized by parameters $\theta$ and evaluates $Q_\theta(s,a)$. In addition, we use two target networks, one for the actor, parameterized by $\phi^\prime$ and one for the critic, parameterized by $\theta^\prime$, i.e., $\pi_{\phi^\prime}(s)$ and $Q_{\theta^\prime}(s,a)$.

In line 7 of algorithm \ref{algRL}, the action to take given the state $s$ is generated as 
\beq \label{ddpg_act}
a \leftarrow \pi_{\phi}(s)+\epsilon
\eeq
where $\epsilon$ is generated from an Uhlenbeck-Ornstein process \citep{PhysRev.36.823} (noise is correlated between each learning step in algorithm \ref{algRL}).
At each learning step in algorithm \ref{algRL}, the critic is updated using the loss function
\beq\label{ddpg_Jloss}
J(\theta)=\|r(s,a)+\gamma Q_{\theta^\prime}\left(s^\prime,\pi_{\phi^\prime}(s^\prime)\right)- Q_\theta(s,a)\|^2
\eeq 
which is quite similar to (\ref{Jloss}) except that the maximization step with respect to $a^\prime$ is replaced by evaluation of the target policy $\pi_{\phi^\prime}(s^\prime)$.
Afterwards, the policy is updated to maximize the Q-value $Q(s,a)$, therefore the loss to be minimized with respect to $\phi$ is given by
\beq \label{ddpg_Jphi}
J(\phi)=-Q_{\theta}\left(s,\pi_{\phi}(s)\right).
\eeq
Using gradient descent, both $\theta$ and $\phi$ are updated as
\beq \label{ddpg_thetaup}
\theta \leftarrow \theta - \mu_Q \nabla_\theta J(\theta),\ \
\phi \leftarrow \phi - \mu_\pi \nabla_\phi J(\phi)
\eeq
where $\mu_Q$ and $\mu_\pi$ are the learning rates for the critic and the actor, respectively. At each learning step, the target network parameters are also updated by a small amount using Polyak averaging as
\beq \label{ddpg_polyak}
\theta^\prime \leftarrow \tau \theta + (1-\tau) \theta^\prime,\ \
\phi^\prime \leftarrow \tau \phi + (1-\tau) \phi^\prime
\eeq
where $\tau$ is a small positive value.
\subsubsection{Twin delayed DDPG (TD3)}
The main shortcoming of DDPG is the overestimation of the Q-value \citep{TD3} and TD3 is an attempt to overcome this. In this algorithm we use two Q-functions instead of one, parameterized by $\theta_1$ and $\theta_2$, i.e., $Q_{\theta_1}(s,a)$ and $Q_{\theta_2}(s,a)$. Similar to DDPG, each Q-function has its corresponding target, parameterized by $\theta_1^\prime$ and $\theta_2^\prime$, i.e., $Q_{\theta_1^\prime}(s,a)$ and $Q_{\theta_2^\prime}(s,a)$. At initialization, $\theta_1$ and $\theta_2$ are randomly initialized to be as different as possible. The actor uses policy $\pi_\phi(s)$ parameterized by $\phi$ and target policy $\pi_{\phi^\prime}(s)$ with $\phi^\prime$ as parameters.

The agent generates an action in line 7 of algorithm \ref{algRL}, given state $s$ as 
\beq \label{td3_act}
a \leftarrow \mathrm{clip}(\pi_{\phi}(s)+\epsilon,\underline{a},\overline{a})
\eeq
where $\epsilon$ is a vector (similar to the size of the action) whose values are zero mean Gaussian noise with $\sigma=0.2$ standard deviation that is clipped to have values in $[-c,c]$ with $c=0.5$ (the range of values from $-c$ to $c$ is denoted as $[-c,c]$). Furthermore, the action itself is clipped to only have values in $[\underline{a},\overline{a}]$. The objective of this clipping is to smoothen the generated actions, so to act as a form of regularization on $\pi_{\phi}(s)$.

The learning step (line 10 in algorithm \ref{algRL}) is as follows. Using the next state $s^\prime$ (sampled from $\mathcal{D}$) and using the target policy, we generate
\beq \label{td3_tact}
a^\prime \leftarrow \mathrm{clip}(\pi_{\phi^\prime}(s^\prime)+\epsilon,\underline{a},\overline{a})
\eeq
where $\epsilon$ and the clipping operations are similar to (\ref{td3_act}). Because we have two Q-functions, the loss for the $i$-th ($i=[1,2]$) can be written as 
\beq \label{td3_qloss}
J(\theta_i)=\|r(s,a)+\gamma \underset{j=[1,2]}{\mathrm {min}}Q_{\theta_j^\prime}\left(s^\prime,a^\prime\right)- Q_{\theta_i}(s,a)\|^2
\eeq
where we find the minimum Q-value of the target Q-functions evaluated at $s^\prime,a^\prime$ by $\underset{j=[1,2]}{\mathrm {min}}Q_{\theta_j^\prime}\left(s^\prime,a^\prime\right)$. By comparing two Q-values and using the minimum of the two, we try to avoid overestimating the Q-value, which is an improvement to DDPG.
We minimize the total loss $J(\theta_1)+J(\theta_2)$ (averaged over the mini-batch) to update both $\theta_1$ and $\theta_2$.

In TD3, the policy is updated less frequently than the Q-functions. While at each learning step $\theta_1$ and $\theta_2$ are updated, the parameters of the policy $\phi$ are updated with a slower cadence (say at every 5 learning steps). The policy update is done by maximizing the Q-value with respect to $\phi$, quite similar to DDPG (\ref{ddpg_Jphi}) except we only use one Q-function for this, for example we minimize
\beq \label{td3_aloss}
J(\phi)=-Q_{\theta_1}\left(s,\pi_{\phi}(s)\right).
\eeq
Notice that during the policy update, we do not add noise to the actions as in (\ref{td3_act}) or (\ref{td3_tact}).

The target parameters are updated similar to DDPG (\ref{ddpg_polyak}) except that it is performed with the same cadence with which the actor is updated, as in
\beq \label{td3_polyak}
\theta_i^\prime \leftarrow \tau \theta_i + (1-\tau) \theta_i^\prime,\ i=[1,2], \
\phi^\prime \leftarrow \tau \phi + (1-\tau) \phi^\prime
\eeq
where $\tau$ is a small positive value.
\subsubsection{Soft actor critic (SAC)}
Both DDPG and TD3 have deterministic policies $\pi(s)$ and therefore to enable some exploration while choosing actions, some form of noise is added to the action as in (\ref{ddpg_act}) or in (\ref{td3_act}). In contrast, the actor in SAC implements a stochastic policy $\pi(a|s)$ that inherently includes some randomness. Each element of vector $a$ is modeled to be an independent and identically distributed random variable to create the policy $\pi(a|s)$. The probability density function (PDF) of the $i$-th element of $a$ (say $a_i$) is generated using a Gaussian random variable with mean $\mu_i$ and variance $\sigma_i^2$. However, this leads to values in the range $[-\infty,\infty]$ and to keep the actions within a finite range, the probability density function is transformed by a $\mathrm{tanh}(\cdot)$ function that maps the output to $[-1,1]$. In practice we use a DNN with parameters $\phi$ to model $\pi_{\phi}(a|s)$ as the mean and the variance of each element $a_i$.

Similar to TD3, SAC uses two Q-functions parameterized by $\theta_1$ and $\theta_2$ as $Q_{\theta_1}(s,a)$ and $Q_{\theta_2}(s,a)$ and corresponding target Q-functions $Q_{\theta^\prime_1}(s,a)$ and $Q_{\theta^\prime_2}(s,a)$ parameterized by $\theta^\prime_1$ and $\theta^\prime_2$. 
In order to generate an action in line 7 of algorithm \ref{algRL}, we sample the policy as
\beq \label{sac_act}
a \sim \pi_{\phi}(\cdot|s).
\eeq

The learning step in line 10 of algorithm \ref{algRL} involves minimizing loss functions to update the critic ($\theta_1$ and $\theta_2$) and thereafter to update the policy ($\phi$). Generally, high randomness in actions leads to high chance of exploration (as opposed to exploitation). High randomness also means high differential entropy of the conditional PDF, so we can increase the reward given to policies with high entropy. With this motivation, we add an extra reward to the Q-value based on the differential entropy of the policy $\pi_{\phi}(a|s)$ that is proportional to $-\log \pi_{\phi}(a|s)$. With this motivation, we have a loss function for the critic as
\beqn \label{sac_qloss}
\lefteqn{J(\theta_i)=}\\\nonumber
&&\hspace{-3em}\|r(s,a)+\gamma \left(\underset{j=[1,2]}{\mathrm {min}}Q_{\theta_j^\prime}\left(s^\prime,a^\prime\right) - \alpha \log\pi_{\phi}(a^\prime| s^\prime)\right)\\\nonumber
&&\mbox{}- Q_{\theta_i}(s,a)\|^2
\eeqn
where
\beq \label{sac_tact}
a^\prime \sim \pi_{\phi}(\cdot|s^\prime)
\eeq
and $\alpha$ is the entropy regularization factor (small value $\approx 0.1$). Note that we need to apply the $\mathrm{tanh}(\cdot)$ to the Gaussian PDF to calculate $\log\pi_{\phi}(a| s)$ in closed form, as given in \cite{SAC,SAC1}.

The policy is updated by minimizing
\beq \label{sac_aloss}
J(\phi)=-\left(\underset{j=[1,2]}{\mathrm {min}}Q_{\theta_j}\left(s,a_\phi\right) - \alpha \log\pi_{\phi}(a_\phi| s) \right)
\eeq
where
\beq \label{sac_act1}
a_\phi \sim \pi_{\phi}(\cdot|s)
\eeq
is a sample drawn from $\pi_{\phi}(\cdot|s)$. In order to have a differentiable (with respect to $\phi$) cost function $J(\phi)$ in (\ref{sac_aloss}), we use the re-parameterized \citep{VAE} generation of $a_\phi$ as
\beq \label{reparm}
a_\phi \leftarrow \mathrm{tanh}(\mu_\phi+\sigma_\phi \odot \xi),\ \xi\sim \mathcal{N}({\bf 0},{\bf I})
\eeq
where $\odot$ is the element-wise product and $\xi$ is randomly generated from $\mathcal{N}({\bf 0},{\bf I})$ which is Gaussian noise with zero mean and unit covariance.
Note that we still have a differentiable mapping from vectors $\mu_\phi$ (mean) and $\sigma_\phi$ (diagonal covariance) to $a_\phi$. Both $\mu_\phi$ and $\sigma_\phi$ are modeled by a DNN with parameters $\phi$.

After updating $\theta_1$ and $\theta_2$ at each learning step, the target network parameters $\theta_1^\prime$ and $\theta_2^\prime$ are also updated by Polyak averaging as in (\ref{td3_polyak}).

\section{Model based reinforcement learning\label{modelbased}}
\begin{figure}[ht]
\begin{minipage}{0.98\linewidth}
\begin{center}
\epsfig{figure=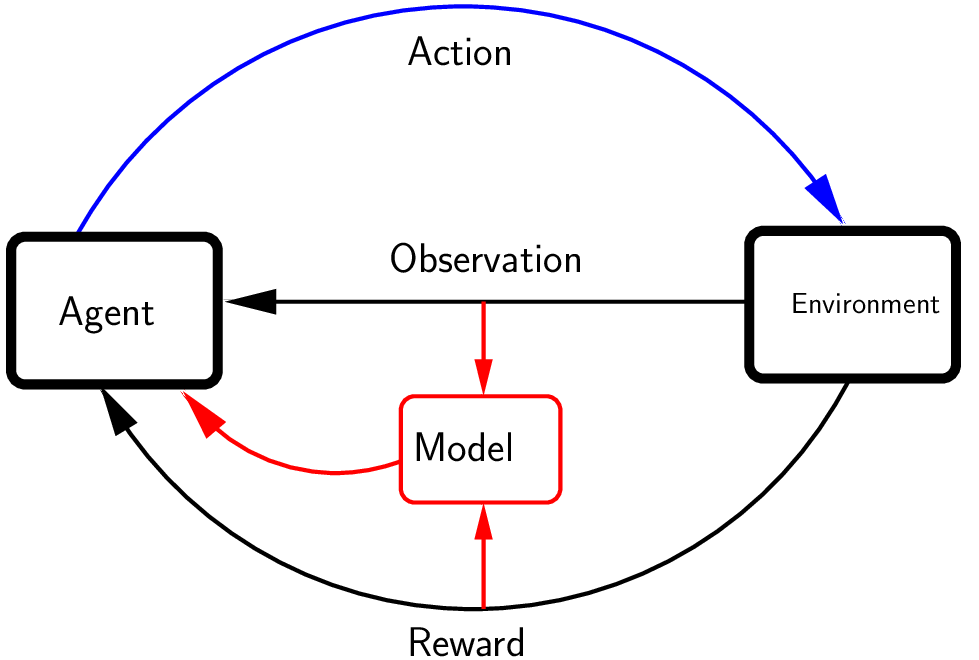,width=8.0cm}\\
\end{center}
\end{minipage}
  \caption{Model based RL. A dynamics model representing the environment is created and used by the agent.\label{model_based}}
\end{figure}

Generating enough data to train RL agents is a fundamental problem in many applications. In some applications, the data generation is expensive and may even harm the actual physical system, for example a robot or an automobile might be damaged if certain actions are performed. In order to overcome this problem, as shown in Fig. \ref{model_based}, we can create a representative model of the actual physical system to generate more data. The use of an internal model to represent the environment is called model based RL and offers a rich variety of algorithms \citep{Levine2015, Anusha2017, Clavera2018, chua2018deep, janner2019trust, Wang2019, Clavera2020, Wang2021}. Of course, model based RL will only work if the model we construct as the proxy for the environment can accurately represent the dynamics of the environment.  

There are two forms of uncertainties that we need to consider when building a model to represent the actual environment \citep{chua2018deep}. Aleatoric uncertainty is the uncertainty due to inherent randomness in the measurements (e.g., thermal noise, quantization). In contrast, epistemic uncertainty is the uncertainty due to the lack of complete information about the actual physical system (e.g., misrepresentation of the state). We can minimize both forms of the aforementioned uncertainties by using an ensemble of probabilistic DNNs as our model. A probabilistic DNN models the transition probability $p(s^\prime|s,a)$ of the next state $s^\prime$ given $s$ and $a$ as opposed to a deterministic DNN which directly gives the next state $s^\prime$ as output. By using a probabilistic DNN we can reduce the aleatoric uncertainty. In order to reduce the epistemic uncertainty, we use an ensemble of such probabilistic DNN models.

\subsection{Probabilistic ensemble models}
We can select any probability distribution to model the transition probability and often a multinomial Gaussian with a mean $\mu_{\theta_i}$ and a diagonal covariance $\Sigma_{\theta_i}$ is chosen, as in
\beq
p_{\theta_i} \sim \mathcal{N}(\mu_{\theta_i},\Sigma_{\theta_i}).
\eeq
We use parameters $\theta_i$ to represent $\mu_{\theta_i}$ and $\Sigma_{\theta_i}$ in the probabilistic model. In order to estimate the parameters $\theta_i$, we maximize the likelihood, or minimize the negative log-likelihood (ignoring the constant terms)
\beqn \label{nll}
\lefteqn{\mathcal{J}({\theta_i})=}\\\nonumber
&&\hspace{-3em}\left(\mu_{\theta_i}(s,a)-s^\prime\right)^T \Sigma^{-1}_{\theta_i}(s,a)\left(\mu_{\theta_i}(s,a)-s^\prime\right)\\\nonumber
&&+ \log \det \Sigma_{\theta_i}(s,a).
\eeqn
We minimize $\mathcal{J}(\theta_i)$ in (\ref{nll}) by using $s,a$ and $s^\prime$ sampled from the replay buffer $\mathcal{D}$ \citep{Ensemble}. In the ensemble, we have $B$ probabilistic DNN models with parameters $\theta_i$, $i\in[1,B]$ for each model. During training, each $\theta_i$ is randomly initialized and we independently sample minibatches with replacement for each $i$ to minimize $\mathcal{J}({\theta_i})$ in (\ref{nll}). The reason for this is to find a diverse set of feasible solutions for $\theta_i$ for each $i$.

Once we have a trained model $p_{\theta_i}$, there are several ways to use it in RL.  A direct way is to use the model to generate more training data \citep{janner2019trust,Wang2021}. In model predictive control \citep{Anusha2017,chua2018deep} on the other hand, the dynamics model is used to predict future rewards and plan the sequence of actions to take. It is also possible to perform gradient descent directly through to model, for example to optimize a policy \citep{Clavera2020}. 

\subsection{Probabilistic ensemble with trajectory sampling}
As an example of model based RL, we discuss the probabilistic ensemble with trajectory sampling algorithm (PETS, \cite{chua2018deep}) in this paper. The PETS algorithm is simpler in the sense that it does not use gradient descent optimization, compared to the model-free algorithms described in section \ref{algo}. However, note that learning the dynamics model by minimizing (\ref{nll}) still involves optimization based on gradient descent. We use an ensemble with $B$ probabilistic models.  The basic idea of PETS is to look ahead $T$ steps and based on the expected reward, decide on which action $a_t$ to take. In algorithm \ref{pets}, we have given the pseudo-code for PETS. Note that we use a replay buffer $\mathcal{D}$ similar to algorithm \ref{algRL}. In fact, algorithm \ref{pets} is similar to \ref{algRL} in the manner of interactions with the environment and the use of a replay buffer. The major difference is given in line 6 of algorithm \ref{pets}, where we select the action $a_t$ by sampling and not by using a policy.
\begin{algorithm}
\caption{PETS \label{pets}}
  \begin{algorithmic}[1]
\REQUIRE Number of episodes $E$, number of loop iterations $L$
\STATE Setup replay buffer $\mathcal{D}$, fill by taking random actions.
\STATE Randomly initialize ensemble dynamics model $p_{\theta_i}$ for all $i$.
  \FOR{$e=1,\ldots,E$}
    \STATE Train $p_{\theta_i}$ using $\mathcal{D}$ by sampling with replacement for all $i$.
    \STATE Select initial state $s_1$ from environment.
    \FOR{$t=1\ldots L$}
    \STATE Run CEM: starting from $s_t$, sample actions for trajectory steps $t,t+1,\ldots t+T-1$ and select optimal action $a_t$. 
    \STATE Execute $a_t$ in the environment, record $s_{t+1}$ and reward $r_t$.
    \STATE $\mathcal{D}$: Store $(s_t,r_t,a_t,s_{t+1})$.
    \ENDFOR
  \ENDFOR
\end{algorithmic}
\end{algorithm}

The selection of the optimal action is done by sampling trajectories starting from the current state $s_t$ using the cross entropy method (CEM, \cite{BOTEV201335}) that is summarized in algorithm \ref{CEM}. We look ahead $T$ steps and let us consider the candidate actions for this trajectory to be $a_t,a_{t+1},\ldots,a_{t+T-1}$. In one trial, we propagate these actions with one (randomly selected) dynamics model $i$ (from $B$ models), as $s_{t+1}\sim p_{\theta_i}(s_t,a_t)$, $s_{t+2}\sim p_{\theta_i}(s_{t+1},a_{t+1})$ and so on until we reach $s_{t+T} \sim p_{\theta_i}(s_{t+T-1},a_{t+T-1})$. Afterwards, using the state and action pairs $(s_t,a_t)$, $(s_{t+1},a_{t+1})$, upto $(s_{t+T-1},a_{t+T-1})$, we calculate the expected rewards $r_t$,$r_{t+1}$, to $r_{t+T-1}$. We roll-out $P$ trials where we randomly select the dynamics model $i$ to generate the state sequence and we record the average reward for this trial as the average reward over all $P$ trials and all $T$ steps.
\begin{algorithm}
\caption{CEM \label{CEM}}
  \begin{algorithmic}[1]
    \REQUIRE State $s_t$, ensemble model $p_{\theta_i}$ $i\in[1,B]$, number of iterations $N$, task horizon length $T$, number of particles $P$, number of candidate actions $C$ and elites $M$
    \REQUIRE Initial mean $\mu_0$ and variance $\sigma_0^2$
    \STATE Setup distribution $\mathcal{N}(\mu,\sigma^2)$ with $\mu\leftarrow \mu_0$ and $\sigma^2\leftarrow \sigma_0^2$.
    \FOR{$n=1,\ldots,N$}
    \FOR{$d=1,\ldots,C$}
    \STATE Sample $a_t,a_{t+1},\ldots a_{t+T-1}$ $\sim \mathcal{N}(\mu,\sigma^2)$
    \STATE Sample $s^p_{t+1}\sim p_{\theta_i}(s_t,a_t)$, $s^p_{t+2}\sim p_{\theta_i}(s^p_{t+1},a_{t+1})$ upto $s^p_{t+T}$ where $i$ randomly selected from $[1,B]$ with replacement and for $p\in[1,P]$.
    \STATE For each $a_\tau$ ($\tau=t,t+1,\ldots,t+T-1$) evaluate $r(s^p_\tau,a_\tau)$ for each $p$.
    \STATE Average $r(s^p_\tau,a_\tau)$ over $p$ ($P$ values) and $\tau$ ($T$ values) and record this for this candidate action.
    \ENDFOR
    \STATE From the candidate actions, find $M$ elite actions corresponding to the $M$ highest averaged rewards.
    \STATE Update $\mu$ and $\sigma^2$ with the mean and variance of the elite actions.
  \ENDFOR
  \STATE Return $\mu$ as optimal action.
\end{algorithmic}
\end{algorithm}

We select $M$ elite actions from the $C$ candidate actions at step $t$, i.e., $a_t$, that has the highest average reward and calculate the mean and variance to update $\mu$ and $\sigma^2$. The updated $\mu$ and $\sigma^2$ are used to generate candidate actions for the next iteration. After $N$ iterations, $\mu$ is returned as the optimal action $a_t$. The crucial component in algorithm \ref{CEM} is the evaluation of the reward (line 6), that needs some information on how the rewards are calculated. If this information is not available, we can train the dynamics model $p_{\theta_i}$ to predict the reward in addition to the state transition probability.
\subsection{Hint assisted RL}
Almost all tasks in astronomy (planning, scheduling, resource allocation, data processing etc.) already have a rich variety of methods in use. Existing methods rely on fundamentals from statistics, signal processing etc. and also on heuristics and experience gained by decades of practice. In such situations, an obvious question to ask is if we can incorporate the knowledge we already have into an RL agent in an efficient manner.

\begin{figure}[ht]
\begin{minipage}{0.98\linewidth}
\begin{center}
\epsfig{figure=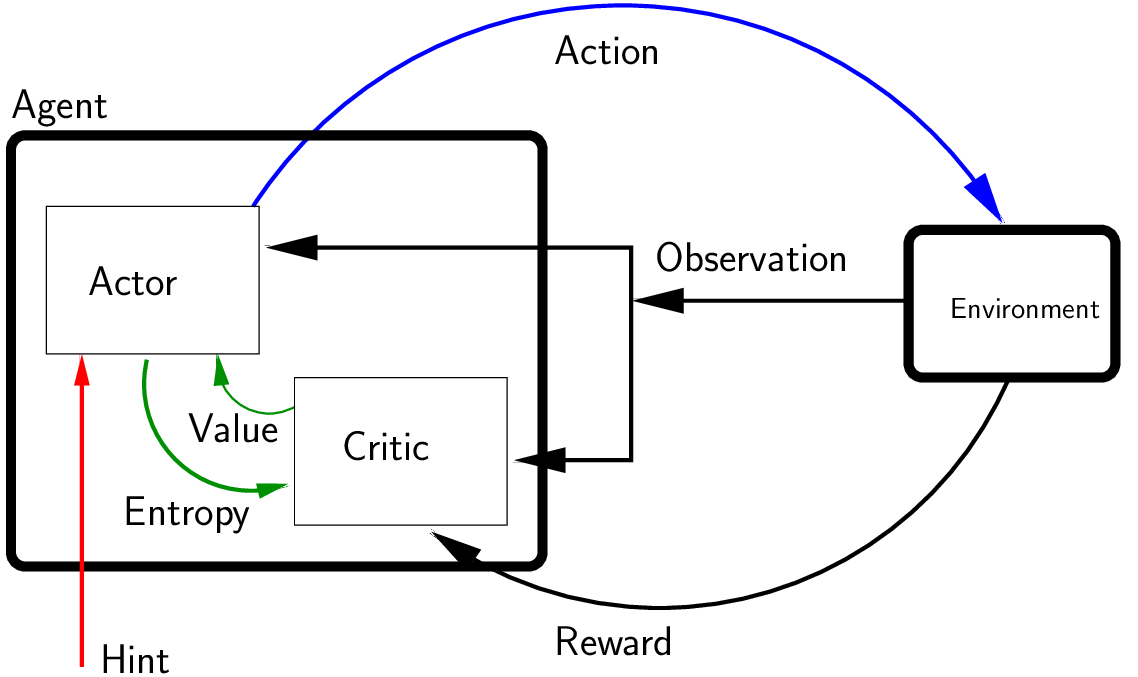,width=8.0cm}\\
\end{center}
\end{minipage}
  \caption{Hint assisted RL. An external hint is directly provided to the actor in the RL agent.\label{hint_assisted}}
\end{figure}

Hint assisted RL \citep{yatawatta2023hint} is one way of incorporating already existing knowledge into the training of RL agents. As shown in Fig. \ref{hint_assisted}, the provision of {\em hints} can be based on anything, for example it could be based on an existing model or it could be based on an experienced astronomer suggesting a solution.

Generally a hint $h$ is a replacement for an action $a$ (having the same dimensionality and the same domain for example). We use a constraint $c(a,h)$ (for example $c(a,h) = \|a-h\|^2$ or any other metric) as a distance measure between the action $a$ and the hint $h$. As shown in Fig. \ref{hint_assisted}, the hint is directly fed into the actor affecting the learning of the policy.

We modify the learning of the policy as follows. First, we define a function $g(a,h)$ as
\beq\label{g}
g(a,h)\buildrel \triangle\over = \left[ c(a,h) - \delta \right]_{+}^2
\eeq
where $\left[x\right]_{+}=x$ if $x>0$ and $\left[x\right]_{+}=0$ otherwise (similar to a ReLU function). In (\ref{g}), a threshold is given by $\delta$ ($>0$) that determines how far apart the hint and the action can be. In other words, we take into account the situations where the hint can be inaccurate, so a large value for $\delta$ indicates that we have less trust in the accuracy of $h$.

We employ the alternating direction method of multipliers (ADMM, \cite{boyd2011,Giesen}) for the policy optimization with the hint being applied as a constraint. We modify (\ref{td3_aloss}) or (\ref{sac_aloss}) as
\beq \label{sac_hint}
J_h(\phi)=J(\phi)
+\frac{\rho}{2}g(a_\phi,h)^2+\lambda g(a_\phi,h)
\eeq
where $J(\phi)$ is the original loss function for the policy in TD3 (\ref{td3_aloss}) or in SAC (\ref{sac_aloss}). The Lagrange multiplier is given by $\lambda$ and $\rho$ is the regularization factor. In the learning step of algorithm \ref{algRL} (line 10), the parameters $\phi$ for the policy are updated as 
\beq \label{phi_up}
\phi \leftarrow \phi - \mu_\pi \nabla_\phi J_h(\phi)
\eeq
where $J_h(\phi)$ is the augmented Lagrangian in (\ref{sac_hint}). Thereafter, the Lagrange multiplier is updated as
\beq \label{lambda_up}
\lambda \leftarrow \lambda + \rho\  g(a_\phi,h).
\eeq
The execution of (\ref{lambda_up}) can be done less frequently than the execution of (\ref{phi_up}). Similar to TD3, the target policy (if exists) parameters can be updated by Polyak averaging as in (\ref{td3_polyak}).
\section{Applications in astronomy\label{app}}
In this section we discuss some practical aspects of applying RL to new tasks and also present some simple examples to illustrate the performance of the algorithms discussed in sections \ref{algo} and \ref{modelbased}. Reinforcement learning algorithms are well supported in both major deep learning frameworks, i.e., Pytorch \citep{Pytorch} and Tensorflow \citep{TF}. Furthermore, an exhaustive number of collections of environments such as Gym \citep{Gym,gymnasium_2023} and collections of standard algorithm implementations such as stable-baselines \citep{stable-baselines} and model based RL libraries \citep{MBRL} also exist.

There are several practical issues that a prospective user of RL should consider when applying aforementioned tools and utilities to their problem.
\begin{itemize}
  \item In some tasks such as in a telescope system control, the state can be obvious and can directly relate to the physical measurements. On the other hand, in more abstract tasks such as in tuning a regression problem, this might not be the case. Therefore some amount of insight into the problem and also some experimentation is required to determine the state representation. An obvious question to ask is if the task behaves as an MDP. If this is not the case and the state transition depends not only on the current state but some history as well, the historical data can also be included in the current state (in other words, the current state is a window of states extending into the past). Similarly, the action can be part of the state and the action learnt by the agent can be the incremental action (or the scaling of the action).
  \item Both the state and the action will be formed by combining information from various sources, in other words combining apples and oranges. We need to pay attention to the numerical stability of the DNNs that are used to represent various models such as the actor or the critic. Ideally, all data should have the same dynamic range for the neural networks to perform well. Therefore, when combining data from different sources, we need to pay special attention to scale or normalize data appropriately.
  \item The calculation of the reward in each task is entirely determined by the objective to achieve and constraints such as differentiability does not apply \citep{tadepalli1996scaling,henderson2018deep}. In practice, actions that deliver good results are boosted by scaling the reward up and conversely, penalties can be added to the reward to discourage undesirable actions. Clipping of rewards is also common in practice \citep{hu2020learning}, mainly for the numerical stability of the DNNs.
  \item In some applications, we will encounter actions that include both continuous and discrete variables, i.e., hybrid action spaces. In such situations, we can model the probabilities of the discrete actions as the policy and draw a number according to the highest probability. For example, if we have $K$ possible discrete actions, we can use a vector of $K$ continuous variables and apply a soft-max operation to get a vector of probabilities. Afterwards, we can choose the element with the highest probability as the discrete action.
  \item The dimensions of the input and the output can vary even within a single problem. For example, the sky models \citep{LSM} used in various data processing steps in radio astronomy can have different number of sources depending on the direction in the sky. In order to accommodate that, we can design our RL agent to handle the largest possible number of dimensions and fill missing values with zeros. An improvement is to use auto-encoders or self attention mechanisms \citep{Vaswani} to encode the input before feeding it into the RL agent. The encoding can be learnt during or before training the RL agent.
\end{itemize}

All forms of tasks in astronomy can be considered as possible applications of RL. We refer the reader back to section \ref{intro} for a review of all such existing applications. We also highlight some of the potential applications of RL in astronomy:
\begin{itemize}
\item Planning and control: Astronomical observatories serve multiple users with diverse science goals. Data collection for such diverse science goals need a significant amount of planning and control and RL can aid in performing this.
\item Resource allocation: Various forms of resources are required to produce science from raw astronomical data. Examples of such resources are: observing duration,  computing resources, storage, network bandwidth etc. The resources are limited and need to be allocated fairly among the users. There are additional constraints such as minimizing the monetary cost and energy consumption. Such allocation problems can also be tackled by RL.
\item Hyper-parameter tuning: Generic pipelines are adapted for processing each specific observation in most astronomical data processing pipelines. In order to do this, various hyper-parameters need to be tuned, mostly by grid-search based approaches. As shown by previous work \citep{Y2021,Ich2021,yatawatta2023hint}, RL agents can outperform grid-search based approaches in tuning tasks such as regression, classification, and clustering.
\item New science: Data collected by various observatories for specific science purposes are stored at large archives of astronomical data. Such archival data can be re-used by other astronomers whose science can be significantly different from the science for which the data were originally collected. We can train RL agents to re-use archival data for new scientific purposes. This discovery can be formulated as a task for an RL agent to learn and the learning can be done in a manner of indexing the archives for potential science.
\end{itemize}

\subsection{Example: Bipedal walker}
\begin{figure}[ht]
\begin{minipage}{0.98\linewidth}
\begin{center}
\epsfig{figure=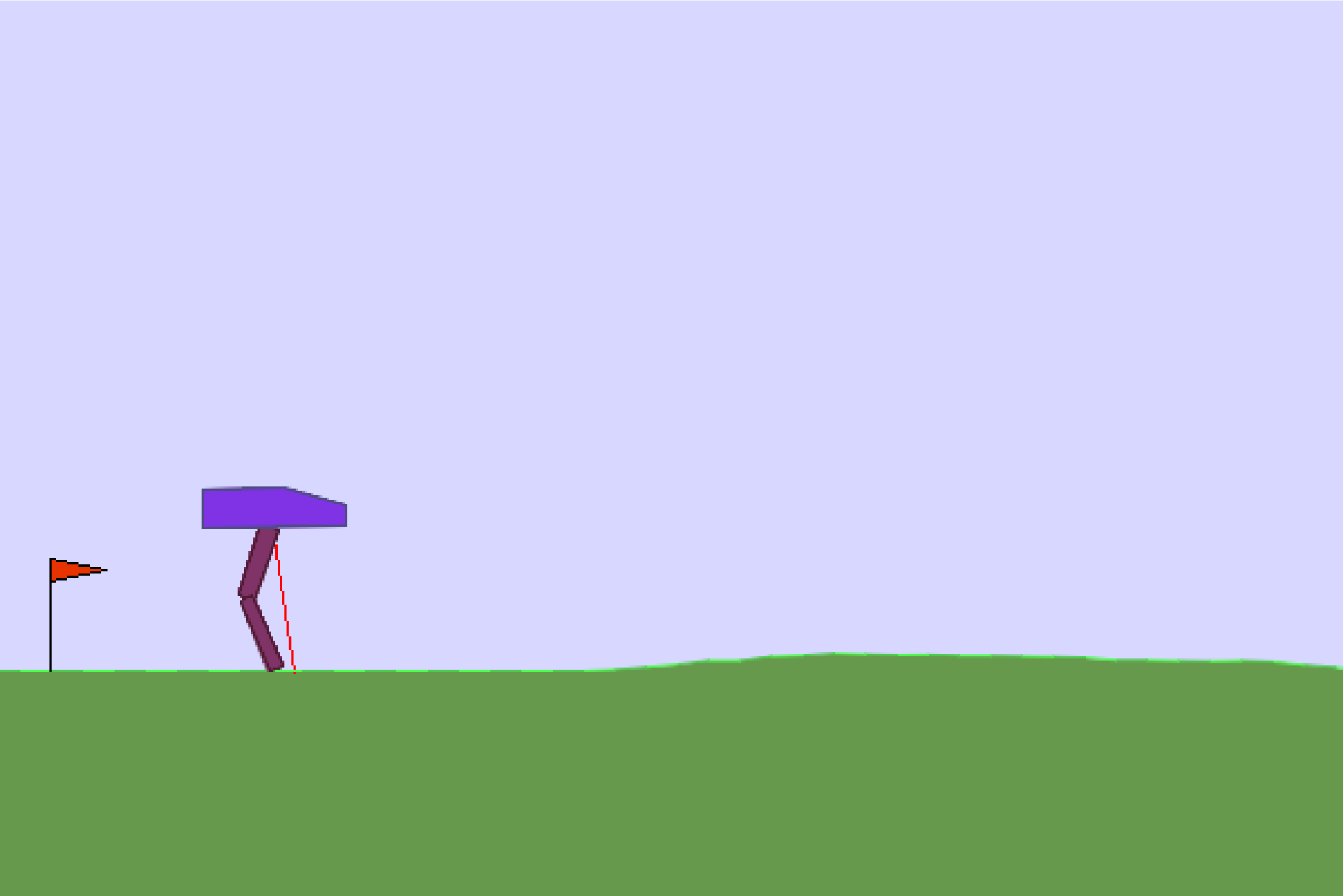,width=8.0cm}\\
\end{center}
\end{minipage}
  \caption{Bipedal walker on flat ground ({\em BipedalWalker-v3}).\label{bp_normal}}
\end{figure}

\begin{figure}[ht]
\begin{minipage}{0.98\linewidth}
\begin{center}
\epsfig{figure=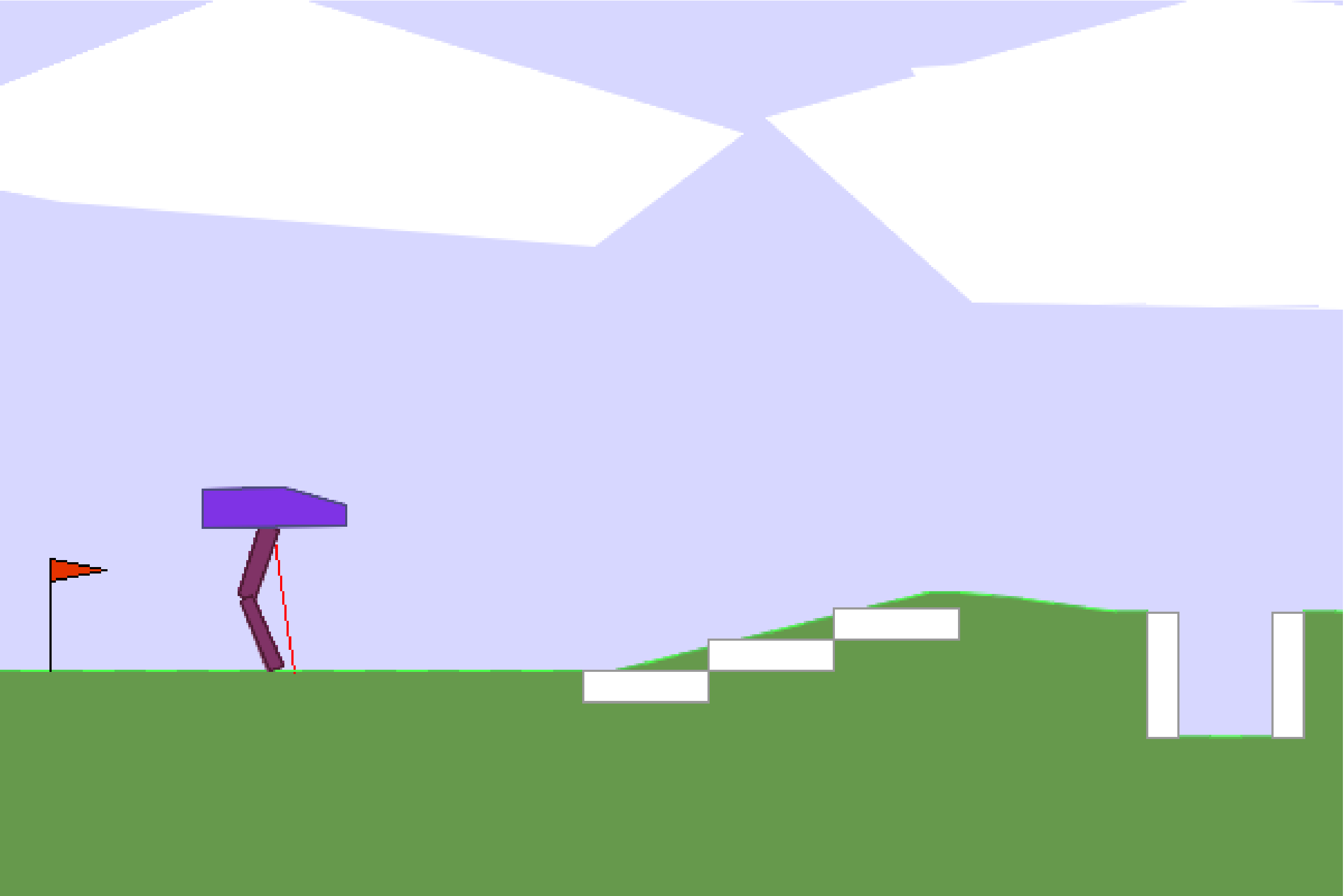,width=8.0cm}\\
\end{center}
\end{minipage}
  \caption{Hard version of bipedal walker with obstacles in the path ({\em BipedalWalkerHardcore-v3}).\label{bp_hard}}
\end{figure}

 Wrapping this section up, we present a simple and yet challenging RL task: training a bipedal robot to walk. This environment is provided with Gym \citep{Gym} and is called the {\em BipedalWalker-v3} environment. The difficulty of training the agent to walk increases depending on the ground on which the robot tries to walk. In Fig. \ref{bp_normal}, we show the environment with almost flat ground that is seemingly a simple task. In contrast, the hard version is shown in Fig. \ref{bp_hard} where the path has many obstacles including stairs, pits and hurdles. 

The state $s$ of the bipedal walker is a vector of $24$ real numbers corresponding to position of leg joints and various velocities of the body. The action $a$ corresponds to the torques applied to the $4$ leg joints, thus the action space is a continuous $4$ dimensional space.

\begin{figure}[ht]
\begin{minipage}{0.98\linewidth}
\begin{center}
\epsfig{figure=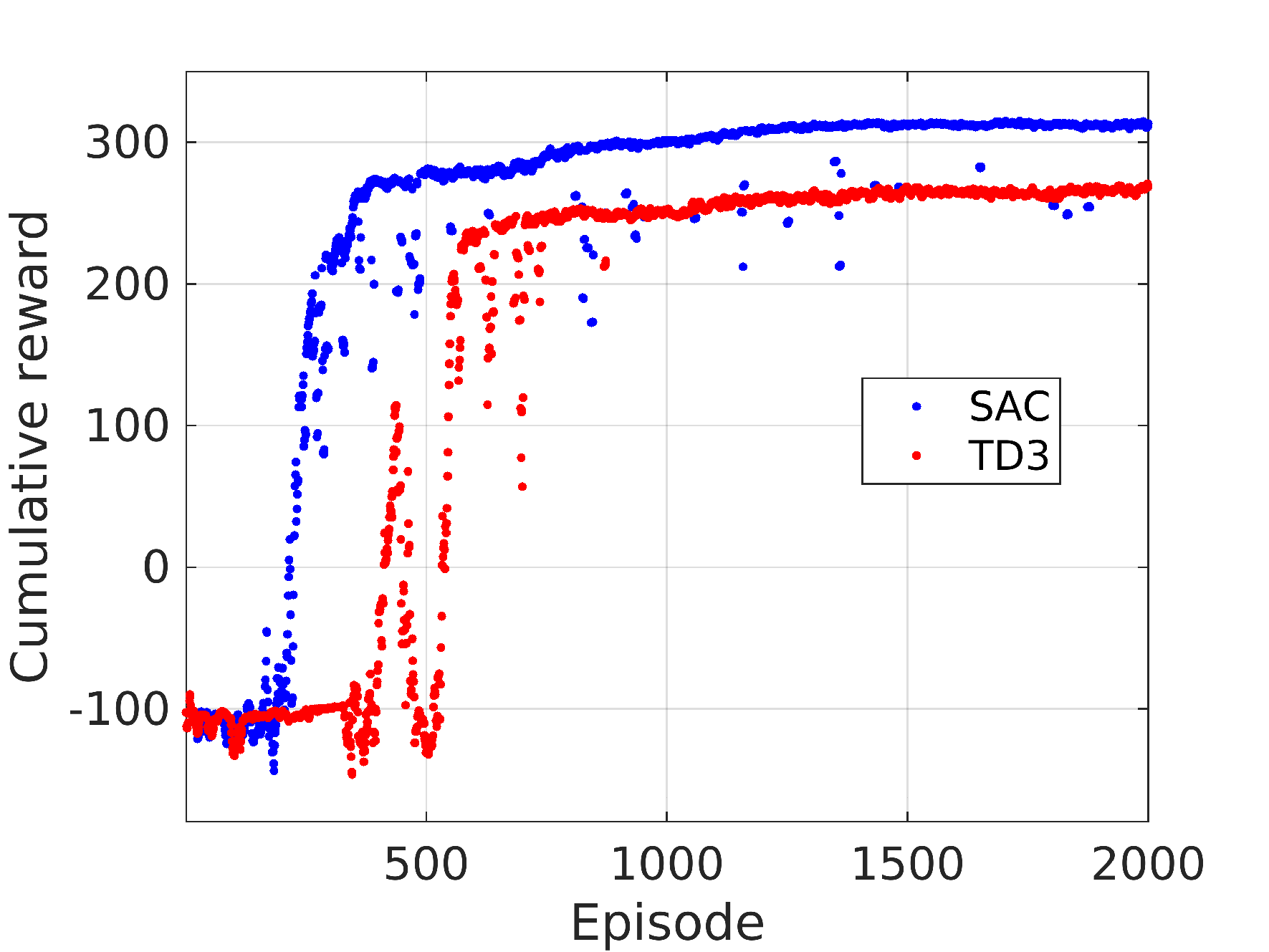,width=8.0cm}\\
\end{center}
\end{minipage}
  \caption{Cumulative reward of the bipedal walker environment with episodes for SAC and TD3.\label{bp_normal_reward}}
\end{figure}

\begin{figure}[ht]
\begin{minipage}{0.98\linewidth}
\begin{center}
\epsfig{figure=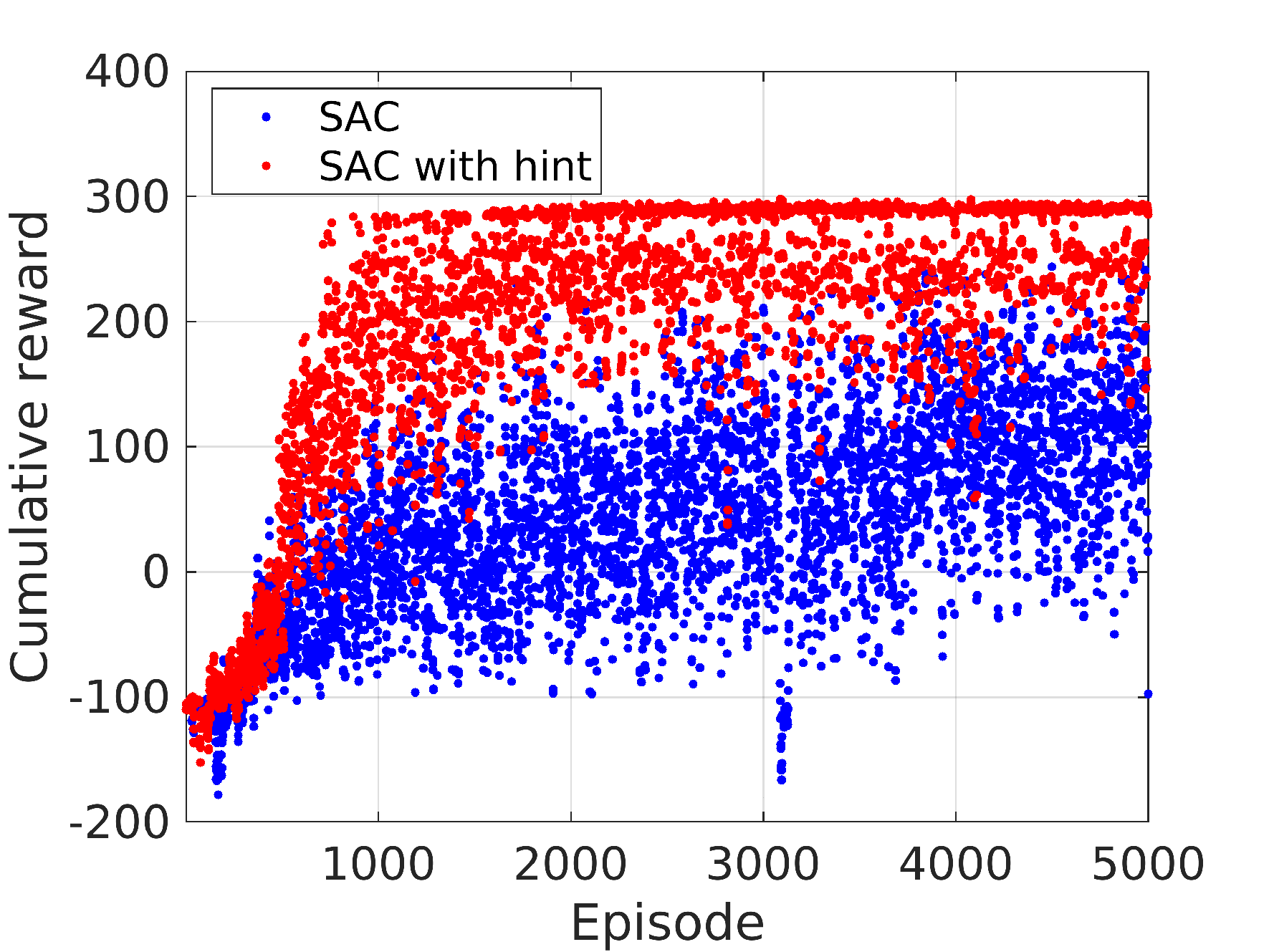,width=8.0cm}\\
\end{center}
\end{minipage}
  \caption{Cumulative reward of the hard version of bipedal walker with episodes for SAC agent with and without using hints.\label{bp_hard_reward}}
\end{figure}

\begin{figure}[ht]
\begin{minipage}{0.98\linewidth}
\begin{center}
\epsfig{figure=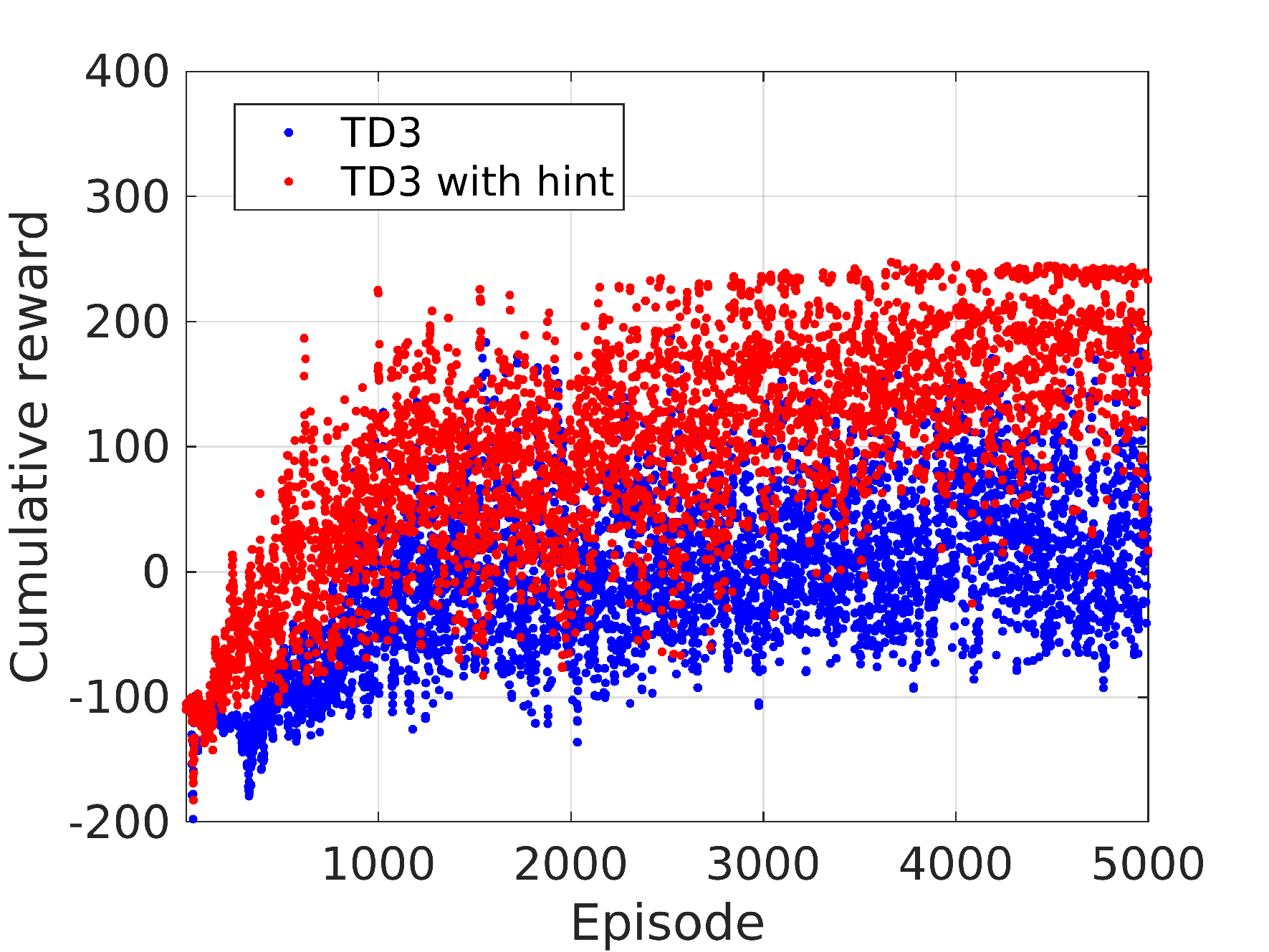,width=8.0cm}\\
\end{center}
\end{minipage}
  \caption{Cumulative reward of the hard version of bipedal walker with episodes for TD3 agent with and without using hints.\label{bp_hard_reward_td3}}
\end{figure}

We employ several algorithms to train both the easy and hard versions of the bipedal walker and show the results in Figs. \ref{bp_normal_reward}, \ref{bp_hard_reward} and \ref{bp_hard_reward_td3}. An agent is considered to have learnt to walk and reach the end of the path once the cumulative reward per episode is more or equal to $300$. The DNN architectures for the actor and the critic and the hyper-parameters used in TD3 and SAC are given in \ref{hyper}. In each comparison, we initialize the environment and the agents using a fixed random seed.

In Fig. \ref{bp_normal_reward} we compare the model free algorithms TD3 and SAC in training the bipedal walker to walk in the easy environment shown in Fig. \ref{bp_normal}. We see that with the SAC algorithm we are able to train the walker to reach the target cumulative reward while with TD3 we get a slightly lower reward. We move on to the hard environment in Figs. \ref{bp_hard_reward} and \ref{bp_hard_reward_td3}. In Fig. \ref{bp_hard_reward}, we compare the performance of the SAC algorithm with and without using hints. The hints in this case are generated by an agent that is trained with the easy environment (Fig. \ref{bp_normal_reward}). Therefore, the hints provided are inherently not accurate but nonetheless, the walker trained with the use of hints is able to achieve a higher reward, almost reaching the target. The terrain in the hard environment has many obstacles such as stairs, pits and hurdles that are randomly generated as seen in Fig. \ref{bp_hard}. This makes the environment highly non-stationary which is also reflected by the dispersion of the rewards achieved in each episode as in Fig. \ref{bp_hard_reward}. This also prevents the vanilla version of SAC (without using hints) from achieving the target reward of $300$.

In Fig. \ref{bp_hard_reward_td3} we perform a similar comparison using the TD3 algorithm. We compare the performance of TD3 with and without using hints in the hard environment. The hint assisted version of TD3 is able to get a higher reward, but much less than the SAC version. The hints are provided by the TD3 agent trained in the easy bipedal walker environment but as we see in Fig. \ref{bp_normal_reward}, the TD3 agent is still not reaching the target reward. Therefore we are using an incompletely trained agent to provide the hints, which explains the poor results in Fig. \ref{bp_hard_reward_td3} compared to Fig. \ref{bp_hard_reward}. Nonetheless, the version using the hints still get a higher reward than the version without hints in Fig. \ref{bp_hard_reward_td3}. We see the use of the hints as a 'brain transplant' from one agent to another and the hints could well be derived from another source.

\subsection{Example: Calibration}
In this example, we consider a problem more applicable to astronomy. The objective of this example is to give a more conceptual overview of formulating the state, action and reward for a general problem and more in-depth details about this particular example can be found in \citep{yatawatta2023hint}. Given a data vector ${y}$, we build a model describing the data as 
\beq \label{ydata}
{y}=\sum_{i=1}^{K} {s}_i({\theta}) +{n}.
\eeq
Problems similar to (\ref{ydata}) can arise in many applications, including regression, model fitting, calibration and so on. In our model, we have $K$ basis functions ${s}_i(\cdot)$ (generally non linear) that are parameterized by the parameter vector ${\theta}$ and our data is corrupted by noise in the vector ${n}$.

\begin{figure}[ht]
\begin{minipage}{0.98\linewidth}
\begin{center}
\epsfig{figure=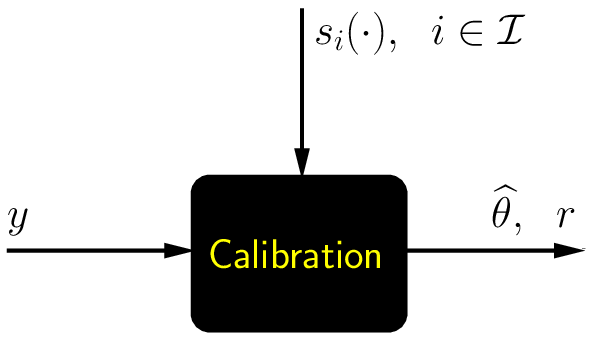,width=8.0cm}\\
\end{center}
\end{minipage}
  \caption{Calibration considered as a black-box.\label{black_box}}
\end{figure}

As illustrated in Fig. \ref{black_box}, we consider the mechanism that actually solves (\ref{ydata}) to be a black-box, i.e., a specialized mechanism to solve the problem at hand of which we do not intend to have low-level control.

The application of RL to the aforementioned problem can be described as follows.
\begin{itemize}
  \item {\em Objective}: We will use RL to determine the best model to use for (\ref{ydata}). We consider the $K$ basis functions ${s}_i(\cdot)$ as our dictionary and we will use RL to select the indices $i$ from $1,\ldots,K$ to best fit our situation. We need to consider two criteria to make this selection. The first criterion is the quality, i.e., we need construct a model that best describes the data. The second criterion is the cost because we have a finite amount of computational resources that limit the size of the model to create as well as the number of iterations that we can expend in creating the model. Generally we will encounter multiple realizations of (\ref{ydata}) with different ground truth models and we use RL to determine the best model for any realization of (\ref{ydata}).
  \item {\em State}: The state should summarize the performance of the system in solving (\ref{ydata}). Since we have a black-box as in Fig. \ref{black_box}, we can use statistical tools such as the influence function \citep{Hampel86,Koh17} to determine the state. The exact details in deriving the influence function for problems similar to (\ref{ydata}) can be found in \citep{ST2019}.
  \item {\em Action}: Let us use the set $\mathcal{I}$ to represent the indices $i$ from $1,\ldots,K$ that we have selected for our model. The action in our problem should specify $\mathcal{I}$ and additionally, the maximum computational budget to expend  within the black-box in Fig. \ref{black_box}. Given $K$ directions, we can formulate the action to predict the probabilities of each $i=1,\ldots,K$ being selected to create $\mathcal{I}$. Therefore, the action can be formulated to be a vector of $K$ values within the range $(0,1]$ and if any value is greater than $0.5$, we select its index to be part of $\mathcal{I}$. In order to determine the computational budget, an additional scalar within the range $(0,1]$ can be appended to the action. We can scale this value to fit within the minimum and maximum computational budget. To recapitulate, the action is formulated to be a vector of $K+1$ values in the range $(0,1]$ derived from the DNN for the actor that can be trained to predict values in the range $[-1,1]$ (or values in the standard normal distribution followed by $\mathrm{tanh}(\cdot)$ activation for a stochastic actor).
  \item {\em Reward}: Once we have $\mathcal{I}$, we determine the parameters $\widehat{\theta}$ and we can find the residual as
\beq \label{residual}
{r}={y}-\sum_{i\in \mathcal{I}}{s}_i(\widehat{\theta}).
\eeq

One way to measure the quality of our model is to use the Akaike information criterion $AIC$ \citep{Akaike}.
Given the standard deviations of the data ${y}$ and the residual ${r}$ as $\sigma_{y}$ and $\sigma_{r}$, respectively, we can determine the $AIC$ as
\beq \label{AIC}
AIC \propto \left(\frac{\sigma_{r}}{\sigma_{y}}\right)^2 \mathrm{length}({y}) + \mathrm{length}({\theta}).
\eeq
The first term of (\ref{AIC}) represents the quality of the model (fractional reduction in the variance by using the model). The second term of (\ref{AIC}) represents the degrees of freedom consumed by the model and $\mathrm{length}({\theta})$ is directly dependent on the cardinality of $\mathcal{I}$ (how many directions are selected).

Using the $AIC$, we can formulate the reward to use for training our RL agent as
\beq \label{reward}
\mathrm{reward}\propto -AIC - \mathrm{penalty}
\eeq
where the $\mathrm{penalty}$ represents the computational cost required to determine the parameters $\widehat{\theta}$ (for example by using maximum likelihood estimation). Since we use an iterative method to find $\widehat{\theta}$, we make the penalty proportional to the number of iterations used by the maximum likelihood estimation algorithm.

\end{itemize}

\begin{figure}[ht]
\begin{minipage}{0.98\linewidth}
\begin{center}
\includegraphics[width=0.9\textwidth]{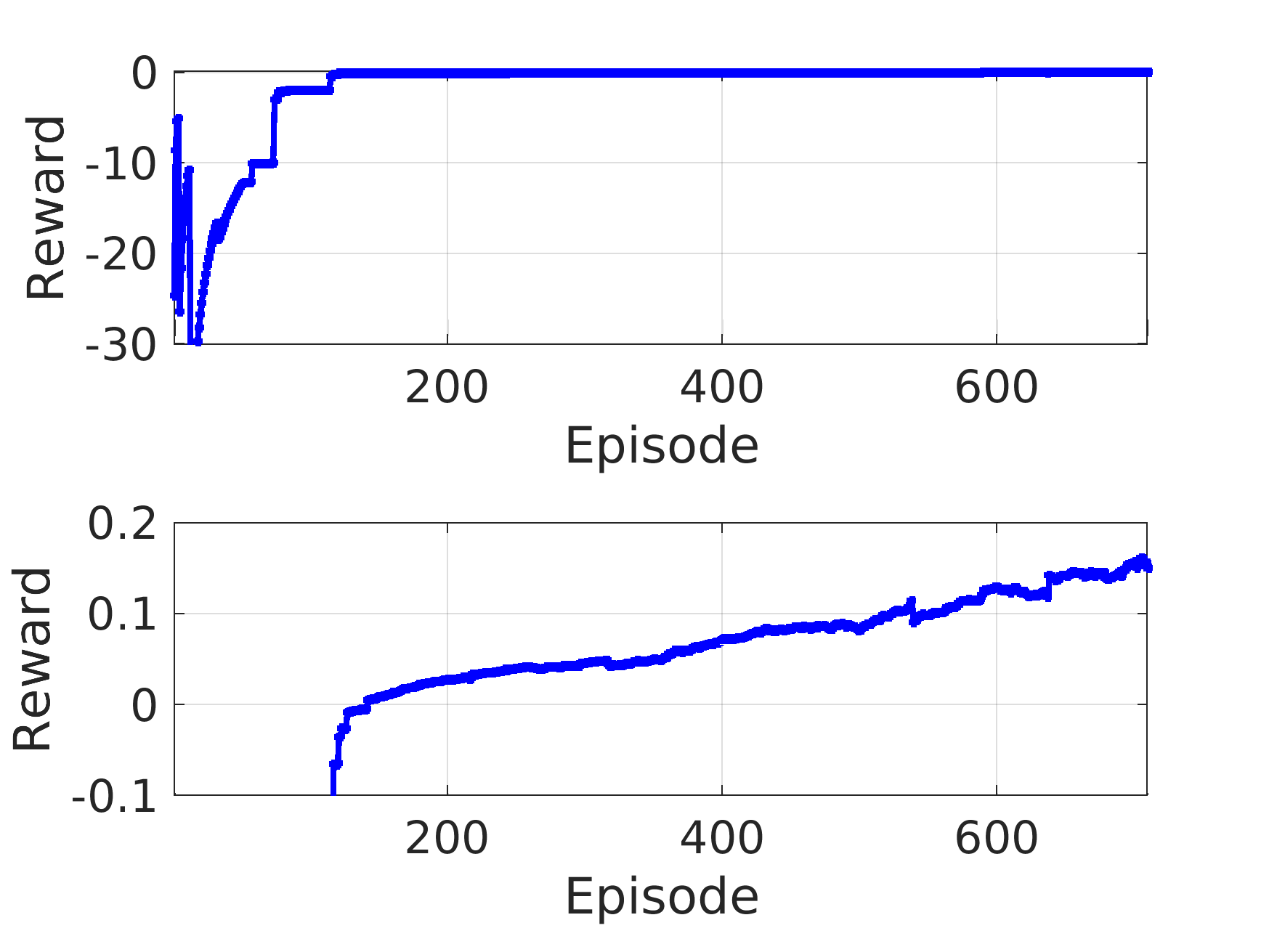}
\end{center}
\end{minipage}
  \caption{Average reward of each episode during learning to solve (\ref{ydata}). The reward is shown in two different scales in the two plots to highlight the early and late behaviors.\label{demix_reward}}
\end{figure}

Guided by the aforementioned concepts, we train an RL agent to maximize the reward for our problem (\ref{ydata}) with $K=6$ using SAC algorithm. Hints are provided to SAC by using an exhaustive search of $2^{K}$ possible choices for $\mathcal{I}$. The midpoint within the minimum and maximum number of iterations is used as the hint for the computational budget. The agent is trained by using simulated random realizations of (\ref{ydata}), each realization is called an episode. In each episode, the agent is given 7 steps to take an action and learn. The evolution of the average reward of each episode is shown in Fig. \ref{demix_reward}. Note that Fig. \ref{demix_reward} shows the same reward curve at two different scales, in order to highlight the initial low rewards and the slow increase of the reward at later iterations showing that the agent is learning.

\section{Conclusions\label{conc}}
We have provided a brief overview of deep reinforcement learning algorithms that are directly applicable in various astronomical tasks. Taking into account the plethora of alternative methods and techniques that already exist in astronomy to perform these tasks, we have also introduced a simple mechanism to transfer the knowledge from existing methods to RL agents via the use of hints. The growth of data intensive astronomy needs efficient and autonomous agents to monitor, control and process data with minimal human involvement. The use of reinforcement learning can help us to achieve this goal.

Source code implementing all algorithms discussed in this paper are publicly accessible at (\href{https://github.com/SarodYatawatta/hintRL}{hint assisted reinforcement learning}).

\section*{Acknowledgments}
We thank Ian Avruch for reviewing an initial version of this paper. We also thank the anonymous reviewer for valuable comments.

\appendix
\section{Python code for Q-table iteration\label{q-python}}
\begin{lstlisting}[language=Python, basicstyle=\tiny, caption=Q-table iteration]
import numpy as np

# Reward
R=np.array([[-1,-np.inf,-np.inf,-1],
       [-1,-np.inf,-1,-np.inf],
       [-1,-1,-np.inf,-1],
       [-np.inf,-1,-1,-np.inf],
       [-np.inf,-1,100,-np.inf]])
# Next state, invalid=-1, terminal=100
Sprime=np.array([[2,-1,-1,1],
       [3,-1,0,-1],
       [4,0,-1,3],
       [-1,1,2,-1],
       [-1,2,100,-1]],dtype=np.int32)

# Discount factor
gamma=0.9
# Q-table
Qtable=np.zeros((5,4),dtype=np.float32)

# function to play one episode
def episode(R,Sprime,gamma,Qt):
 t=0
 s=np.random.choice(np.arange(5))
 while t<100:
  a=np.random.choice(np.argwhere(np.max(R[s])\
    ==R[s]).flatten())
  sprime=Sprime[s,a]
  if sprime==-1:
     print('Error, invalid next state')
     break
  if sprime==100:
     Qt[s,a]=R[s,a]
     break

  print(f'{t}: state {s} action {a} next {sprime}')
  Qt[s,a]=R[s,a]+gamma*(np.max(Qt[sprime]))

  s=sprime
  t=t+1


print(Qtable)
for epoch in range(100):
 episode(R,Sprime,gamma,Qtable)
print(Qtable)
\end{lstlisting}

\section{Hyperparameters in TD3 and SAC\label{hyper}}
The actor and critic network architectures used are as follows:
\begin{itemize}
\item Critic: In both TD3 and SAC, we use a DNN with 3 linear layers. The dimensions of each layer are: input layer $(24+4)\times 256$, hidden layer $256\times 256$, output layer $256 \times 1$.
\item Actor: In both TD3 and SAC, we use a DNN with 3 linear layers, except in SAC the output layer is divided into two heads (for $\mu_\phi$ and $\log \sigma_\phi$). The dimensions of each layer are: input layer $24 \times 256$, hidden layer $256 \times 256$, output layer $256 \times 4$.
\end{itemize}
In all layers except the last, we use ReLU activation. We use the Adam \citep{Adam} gradient descent optimizer in training.

\begin{table}[htbp]
\begin{minipage}{0.98\linewidth}
\caption{Hyper-parameters used in training TD3 and SAC agents.} \label{hyper_tab}
  \begin{center}
   \begin{tabular}{l|l}
     Parameter & Value\\\hline
     Discount $\gamma$ & 0.99 \\
     Batch size & 256 \\
     Learning rate $\mu_\theta$,$\mu_\phi$ & $1e-4$\\
     Polyak averaging $\tau$ & 0.005\\
     Temperature $\alpha$ & 0.036 \\
     ADMM $\rho$ & 0.001\\
     Hint threshold $\delta$ & 0.5\\
   \end{tabular}
  \end{center}
\end{minipage}
\end{table}




\bibliographystyle{model2-names} 
\bibliography{references}





\end{document}